\documentclass[aps,prb,twocolumn,showpacs,floatfix,letterpaper,10pt]{revtex4-1}
\usepackage[intlimits,sumlimits]{amsmath}
\usepackage{amsfonts,amssymb,color}
\usepackage{graphicx}

\pdfoutput=1

\font\elevenmib=cmmib10 scaled 1095
\font\tenmib=cmmib10
\font\eightmib=cmmib10 scaled 800
\font\sixmib=cmmib10 scaled 667
\skewchar\elevenmib='177
\newfam\mibfam

\def\mib{\fam\mibfam\tenmib}
\textfont\mibfam=\tenmib
\scriptfont\mibfam=\eightmib
\scriptscriptfont\mibfam=\sixmib
\mathchardef\alpha="710B
\mathchardef\beta="710C
\mathchardef\gamma="710D
\mathchardef\delta="710E
\mathchardef\epsilon="710F
\mathchardef\zeta="7110
\mathchardef\eta="7111
\mathchardef\theta="7112
\mathchardef\kappa="7114
\mathchardef\lambda="7115
\mathchardef\mu="7116
\mathchardef\nu="7117
\mathchardef\xi="7118
\mathchardef\pi="7119
\mathchardef\rho="711A
\mathchardef\sigma="711B
\mathchardef\tau="711C
\mathchardef\phi="711E
\mathchardef\chi="711F
\mathchardef\psi="7120
\mathchardef\omega="7121
\mathchardef\varepsilon="7122
\mathchardef\vartheta="7123
\mathchardef\varrho="7125
\mathchardef\varphi="7127
\mathchardef\sGamma="7100
\mathchardef\sDelta="7101
\mathchardef\sTheta="7102
\mathchardef\sLambda="7103
\mathchardef\sXi="7104
\mathchardef\sPi="7105
\mathchardef\sSigma="7106
\mathchardef\sUpsilon="7107
\mathchardef\sPhi="7108
\mathchardef\sPsi="7109
\mathchardef\sOmega="710A

\def\gtwid{\,{\raise.3ex\hbox{$>$\kern-.75em\lower1ex\hbox{$\sim$}}}\,}
\def\ltwid{\,{\raise.3ex\hbox{$<$\kern-.75em\lower1ex\hbox{$\sim$}}}\,}

\def\Dpr{{\Delta^{\prime}}} 
\def\bra#1{\left\langle #1 \right|} 
\def\ket#1{\left| #1 \right\rangle} 
\def\braket#1#2{\left\langle\left. #1 \right| #2\right\rangle} 
\def\expect#1#2#3{\left\langle #1 \left| #2 \right| #3 \right\rangle}
\def\half{\frac{1}{2}}

\def\ns{^{\vphantom{*}}}

\def\util{\tilde{u}}
\def\vtil{\tilde{v}}
\def\Bk{{\mib k}}
\def\BK{{\mib K}}
\def\Ba{{\mib a}}

\def\Bq{{\mib q}}
\def\CB{{\cal B}}
\def\CH{{\cal H}}
\def\CE{{\cal E}}
\def\Bvg{{\mib v}\ns_{\rm g}}
\def\vgr{{v\ns_{\rm g}}}
\def\BCB{{\boldsymbol{\cal B}}}
\def\ve{\varepsilon}
\def\vphi{\varphi}
\def\kstar{k\ns_\star}
\def\Estar{E\ns_\star}
\def\vestar{\varepsilon\ns_\star}

\def\Dtil{{\widetilde\Delta}}
\def\Htil{{\widetilde H}}
\def\Etil{{\widetilde E}}
\def\CEtil{{\widetilde {\cal E}}}
\def\ktil{{\widetilde k}}
\def\yd{^\dagger}
\def\ellB{{\ell\ns_{\!B}}}

\def\PhB{\Phi\ns_{\rm B}}
\def\Phc{\Phi\ns_{\rm c}}
\def\sigH{\sigma\ns_{\rm H}}

\def\gze{\gamma\ns_0}
\def\gon{\gamma\ns_1}
\def\gth{\gamma\ns_3}
\def\gfo{\gamma\ns_4}

\def\efo{\eta\ns_4}

\def\Sq{S\ns_\Bq}
\def\Sqs{S^*_\Bq}
\def\xhat{\hat{\mib x}}
\def\yhat{\hat{\mib y}}
\def\zhat{\hat{\mib z}}
\def\omo{\omega\ns_0}
\def\elB{\ell\ns_B}
\def\CH{{\cal H}}
\def\CHp{{\cal H}^+}
\def\CHm{{\cal H}^-}
\def\Bnab{\boldsymbol{\nabla}\!}

\providecommand\Km{K^{-}} %
\providecommand\Kp{K^{+}} %
\DeclareMathOperator{\sign}{sgn}  % sign

\begin{document}

\title{Magnetoelectric coupling, Berry phase, and Landau level dispersion in a biased bilayer graphene}

\author{Lingfeng M. Zhang}
\author{Michael M. Fogler}
\author{Daniel P. Arovas}
\affiliation{University of California San Diego, 9500 Gilman Drive, La
Jolla, California 92093}
\date{\today}
\begin{abstract}

We study the energy spectrum of a graphene bilayer in the presence of transverse electric and magnetic fields. 
We find that the resulting Landau levels exhibit a nonmonotonic dependence on the electric field, as well as numerous level crossings.
This behavior is explained using quasiclassical quantization rules that properly take into account the pseudospin of the quasiparticles.
The pseudospin generates the Berry phase, which leads to a shift in energy quantization and results in a pseudo-Zeeman effect.
The latter depends on the electric field, alternates in sign among the two valleys, and also reduces the band gap.
Analytic formulas for other pseudospin-related quantities, such as the anomalous Hall conductivity, are derived and compared with prior theoretical work.

\end{abstract}


\maketitle
\section{Introduction}
\label{sec:Introduction}

%%%%%%%%%%%%%%%%%%%%%%%%%%%%%%%%%%%%%%%%%%%%%%%%%%%%%%%%%%%%%%%%%%%%%%%%%%%%%%%

\begin{figure}[b]
  \centering
  \includegraphics[width=3.2in]{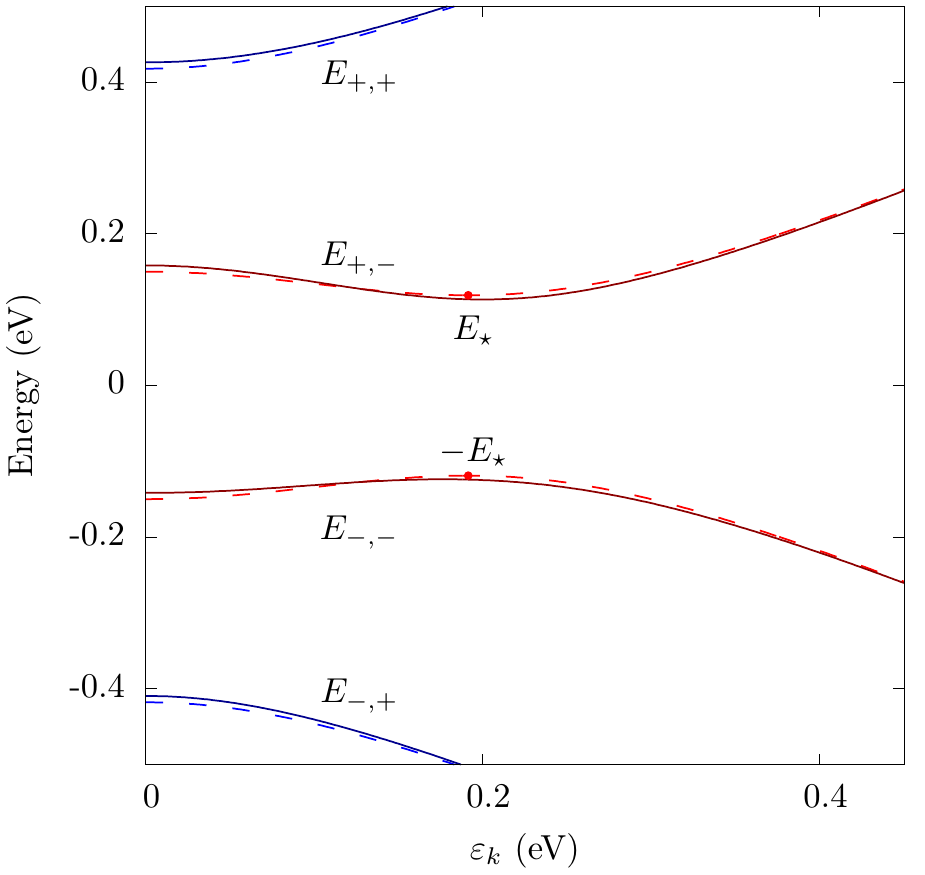}
  \caption{BLG band dispersion as a function of $\ve_k = \hbar v_0 k$, where $k$ is the momentum measured from the nearest $K^\pm$ point.
At $B = 0$ the bands are valley-degenerate.
The dashed curves show their dispersion calculated from Eq.~\eqref{eqn:BLG_E_k} for the interlayer bias $2U = 240\,\textrm{meV}$.
In a finite field, the bands acquire a pseudo-Zeeman shift, Eq.~\eqref{eqn:E_shifted}, opposite in the two valleys.
The solid curves show the result for $K^+$ at $B = 5\,\text{T}$.
  \label{fig:BLG_E_bands}}
\end{figure}

The physics of monolayer and bilayer graphene has attracted much recent attention.~\cite{Castroneto2009tep} 
A unique feature of bilayer graphene (BLG) is its tunable band structure: the symmetric bilayer is gapless but when an interlayer potential difference $U$ is induced, a band gap opens.
The low-energy regions affected by the gap are situated at the Brillouin zone corners, e.g., points ${\mib K}^\pm = \pm ({4\pi} / {3a\ns_0}) \xhat$ henceforth referred to as $K^\pm$ valleys, near which the band dispersion acquires a ``Mexican hat'' shape~\cite{McCann2006lld, Nilsson2006epo} (see Fig.~\ref{fig:BLG_E_bands}), where $a\ns_0 = 2.46\,\text{\AA}$ is the lattice constant for the underlying triangular Bravais lattice.

There have been interesting theoretical predictions that electron interactions can spontaneously generate layer polarization and a band gap,~\cite{Nilsson2006eei, Min2008pmi, ZhangF2009eei, Nandkishore2010dsa, Vafek2010mbi, Lemonik2010ssb} but they have yet to be verified experimentally.~\footnote{Some encouraging results in this direction have been reported in Refs.~\onlinecite{Feldman2009bss}.
However, the observed gap is comparable to that in the single-particle picture, see Fig.~\ref{fig:BLG_LL_Vb_zoom} in Sec.~\ref{sec:Prior}.
The gaps measured in another experiment~\cite{Zhao2010sbi} are even smaller, possibly due to much stronger disorder.} The proven ways of creating an interlayer bias $U$ include doping and gating.
The latter enables one to change $U$ continuously, although the dependence of $U$ on the gate voltage is nontrivial.~\cite{McCann2006agi, Castro2007bbg}
In most of experimental studies of bilayer graphene a single gate electrode was used.~\cite{ZhangY2005eoo, Novoselov2006uqh, Ohta2006ces, Castro2007bbg, Henriksen2008cri, Li2008dcd, Mak2009ooa, Kuzmenko2009iso, Kuzmenko2009dot, Kuzmenko2009gti, Zhao2010sbi}
In such devices the interlayer bias $U$ and the induced electron density $n$ vary concomitantly with the gate voltage.
Separate control of $U$ and $n$ can be achieved with two gates.~\cite{ZhangY2009doo, Fogler2010sig}  Experiments with dual-gate devices~\cite{Oostinga2007gii, ZhangY2009doo, Kim2009qhe, Feldman2009bss, Henriksen2010mot} have been reported recently.

Another intriguing property of graphene is that its low-energy quasiparticles are endowed with a pseudospin-$\half$ degree of freedom, associated with the sublattice structure of each monolayer, whose dynamics is linked to their orbital motion.~\cite{Castroneto2009tep}
When a quasiparticle traces a closed-loop trajectory in momentum space, its pseudospin sweeps out a certain solid angle, just as in the canonical
Berry phase setting.~\cite{Berry1984qpi, Shapere1989gpp}
Such orbits naturally occur when an external magnetic field $B$ is present --- they are the cyclotron orbits.
In monolayer graphene the corresponding Berry phase is equal to $\pi = \half \times (2 \pi)$ at all energies.~\cite{Mikitik1999mob}
This property is the reason for the $\half$-shift in the Landau level filling factor $\nu = 4 \times \big(N - \half \big)$ at which $N${th} magnetoresistance minimum occurs.~\cite{Zheng02hco, Gusynin2005uiq}
Here the factor of four is the spin-valley degeneracy, assuming it is preserved.

Given the unusual band structure of BLG, it is interesting to consider the effects of the Berry phase and other pseudospin-related phenomena on the Landau levels and the magnetic response in this material.
Indeed, it is known~\cite{Xiao2007vpg, Koshino2010aom} that the pseudospin generates a linear coupling to the transverse component $B_z$ of the magnetic field, similar to a real spin.

Note that such a pseudo-Zeeman coupling does not violate the time reversal symmetry of the system at $B = 0$.
Since this symmetry operation interchanges the valleys, it is only the sum $M_z^+ + M_z^-$ of the corresponding magnetic moments that must vanish.
Further symmetry considerations require the pseudo-Zeeman shift of the energy eigenvalue $E\ns_\Bq$ to be linear in both applied fields,
\begin{equation}
\Delta E\ns_\Bq \propto - E_z B_z \cos 3\phi\ns_\Bq \,,
\label{eqn:pseudo-Zeeman}
\end{equation}
where $\phi\ns_\Bq$ is the polar angle in reciprocal space relative to the zone center $\Bq = 0$.
This expression conforms to the following valley-interchanging operations: (i) a reflection $O\ns_1$ with respect to the $y$-$z$ plane, and
(ii) a composite operation $O\ns_2$ consisting of a rotation through angle $\pi$ around $x$-axis in the midplane, followed by time reversal.
Both of these operations leave the crystal structure invariant (see Fig.~\ref{fig:bilayer}).
The first one keeps $E_z$ the same but reverses the sign of $B_z$ (because ${\mib B}$ is a pseudovector).
The second changes the sign of $E_z$ but keeps $B_z$ the same.  

Equation~\eqref{eqn:pseudo-Zeeman} constitutes a magnetoelectric effect in bilayer graphene.
It implies that the valley symmetry cannot be broken solely by $B_z$ or by $E_z$ alone.
Rather, both fields must be nonzero simultaneously.
(It is also reminiscent of the Chern-Simons term which occurs in topological insulators.~\cite{Qi2008tft})
Below we study this kind of valley-symmetry breaking analytically, focusing on the question how it modifies the Landau level dispersion.

Prior theoretical studies~\cite{McCann2006lld} have already showed that Landau levels in bilayer graphene become valley split at finite $U$.
This was explained by noting that the quasiparticle wavefunctions of the two valleys have different dipole moments in the $z$-direction.
Equation~\eqref{eqn:pseudo-Zeeman} offers a complementary interpretation: the two valleys in a biased bilayer graphene have different magnetic moments.~\cite{Xiao2007vpg, Koshino2010aom}

The ratio of the pseudo-Zeeman term~\eqref{eqn:pseudo-Zeeman} and the Zeeman energy due to real spin determine the effective $g$-factor of bilayer graphene.
We show below that $g$ can be an order of magnitude higher than its bare value $g = 2$.
This resembles the situation in Bi, another low band gap material.
In fact, there is a mathematical similarity of the low-energy theories~\cite{Falkovsky1966qco} of the two materials.
(Of course, Bi is three-dimensional.)

The dependence of Landau level energies in bilayer graphene on $B_z$ and $U$ is known to be quite complicated
(see, e.g., Refs.~\onlinecite{Mucha-Kruczynski2009tio, Mucha-Kruczynski2009spo, Koshino2010pav}).
We show that it can be understood if one applies quasiclassical quantization to the Mexican hat band structure.
This procedure requires calculating the phase shifts $\Phc$ acquired by quasiparticles on their cyclotron orbits.
Both the pseudo-Zeeman term and the Berry phase contribute to $\Phc$.
As a result, $\Phc$ generally is not an integer multiple of the monolayer value $\pi$. When it does become equal to $\pi$, at certain values of $U$, an interesting phenomenon occurs: adjacent Landau levels of opposite valleys become degenerate.
Therefore, there are an infinite number of Landau level crossings within the same band.

Landau level crossings in the two dimensional electron gas (2DEG) has previously attracted much theoretical~\cite{Kallin1984efa, Falko1993ccs, Jungwirth1998mai, Jungwirth2000pac, Chalker2002qhf, Wang2002mea, Rezayi2003eds} and experimental~\cite{Koch1993sll, Cho1998hst, Piazza1999fop, DePoortere2000rsa, Eom2000qhf, Zeitler2001mai, dePoortere2003cri, Pan2001hat, ZhangX2005mpw, ZhangX2006lla, Gusynin2006tod, Fischer2007teo} interest because the 2DEG then exhibits many of the properties found in ferromagnets.
Therefore, BLG may be a promising system for studying quantum Hall ferromagnetism.

The remainder of this article is organized as follows.
A brief summary of BLG band structure properties is given in Sec.~\ref{sec:Prior}.
The quasiclassical approximation is discussed in Sec.~\ref{sec:quasiclassical}.
Illustrative Landau level spectra are presented in Sec.~\ref{sec:LL_spectrum}.
The anomalous Hall conductivity of the BLG is computed in Sec.~\ref{sec:AHE}.
Concluding remarks are given in Sec.~\ref{sec:conclusion}.
Technical notes are gathered in the Appendix.

%%%%%%%%%%%%%%%%%%%%%%%%%%%%%%%%%%%%%%%%%%%%%%%%%%%%%%%%%%%%%%%%%%%%%%%%%%%%%%%
\section{Analytic results from prior work}
\label{sec:Prior}
\subsection{Zero magnetic field}
\label{sec:zero_field_band}
%%%%%%%%%%%%%%%%%%%%%%%%%%%%%%%%%%%%%%%%%%%%%%%%%%%%%%%%%%%%%%%%%%%%%%%%%%%%%%%
The band structure of BLG, well known from previous literature,~\cite{Castroneto2009tep} is shown in Fig.~\ref{fig:BLG_E_bands}.
In this section we summarize its main properties, focusing on analytic results.

\begin{figure}[t]
\begin{center}
\includegraphics[width=3.3in]{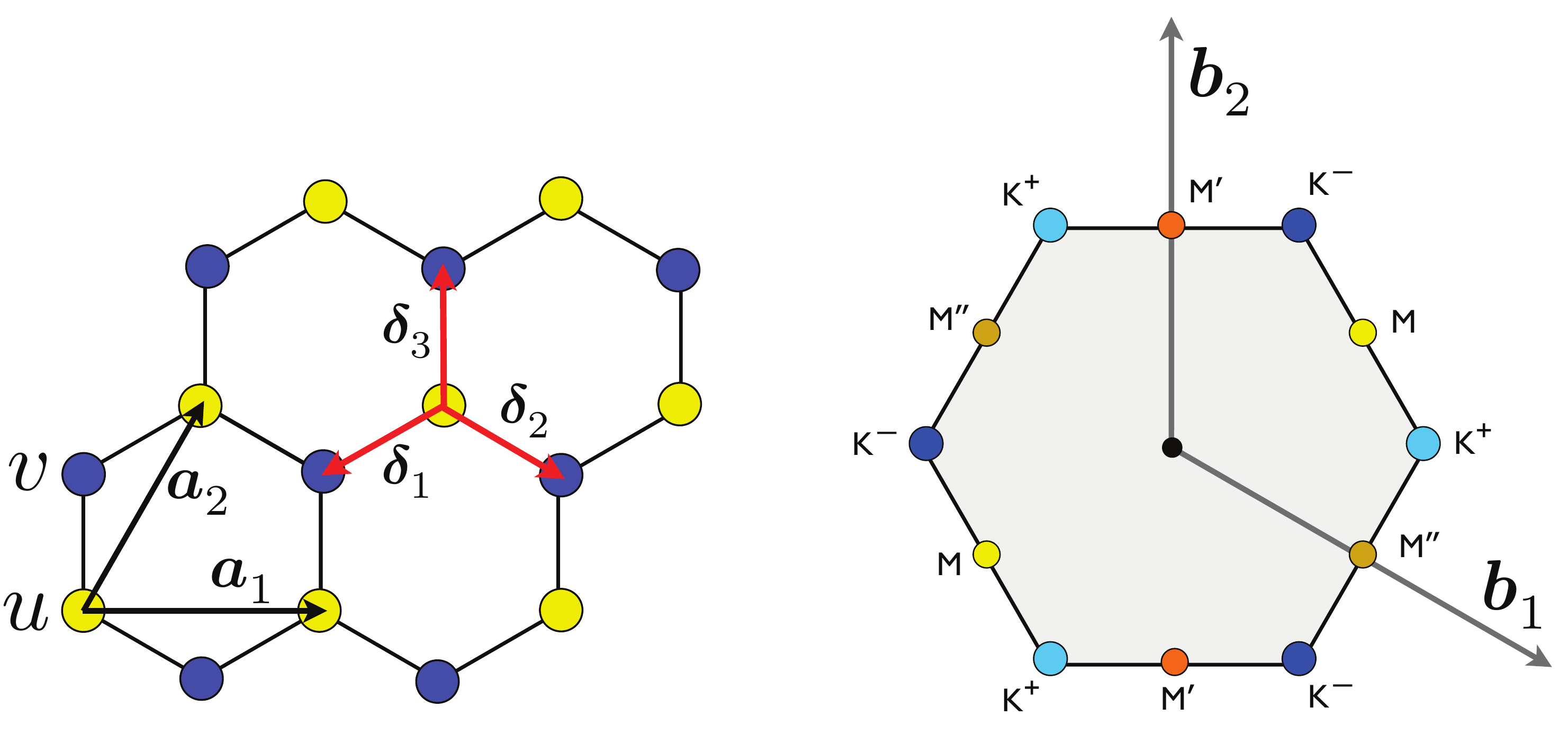}
\end{center}
\caption{(Color online) Graphene monolayer (left) and resulting Brillouin zone (right).}
\label{fig:monolayer}
\end{figure}

\begin{figure}[b]
  \centering
  \includegraphics[width=3.0in]{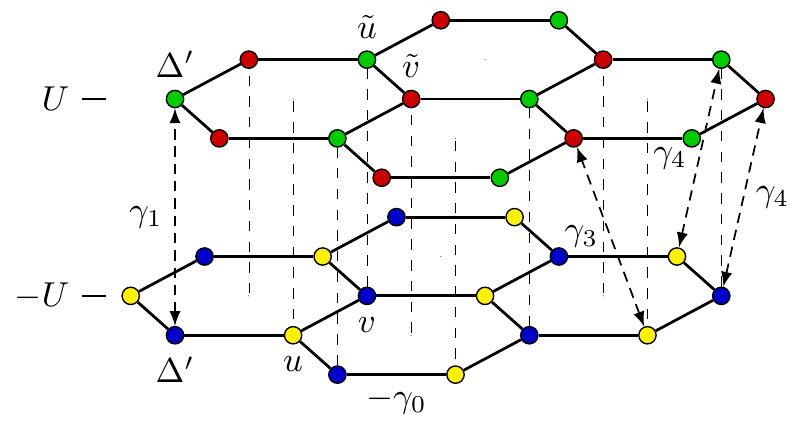}
  \caption{(Color online) Crystal structure of bilayer graphene.
  We label four sublattices by $u$, $v$, $\util$, $\vtil$.
  Also shown is the assignment of the hopping parameters $\gamma_j$ of the tight-binding model.
  The labels $\pm U$ indicate the electrostatic potential energies of the layers.
  \label{fig:bilayer}}
\end{figure}

The unit cell of a graphene bilayer, depicted in Fig.~\ref{fig:bilayer}, consists of four atoms, which we label $u$, $v$, $\util$, and $\vtil$.
The underlying Bravais lattice is the triangular Bravais lattice of either honeycomb monolayer (Fig.~\ref{fig:monolayer}).
The Bravais lattice sites are at locations ${\mib R} = n\ns_1\,\Ba\ns_1 + n\ns_2\,\Ba\ns_2$, where $\Ba\ns_1 = a\ns_0\,\xhat$ and $\Ba\ns_2 = a\ns_0\big(\half\xhat+
\frac{\sqrt{3}}{2}\yhat\big)$ are primitive direct lattice vectors, $n\ns_{1,2}$ are integers, and  $a\ns_0 = 2.461\,$\AA\ is again the lattice constant.
The corresponding elementary reciprocal lattice vectors are ${\mib b}\ns_1 = \frac{4\pi}{a\ns_0\sqrt{3}} \big(\frac{\sqrt{3}}{2} \xhat - \half\yhat\big)$ and ${\mib b}\ns_2 = \frac{4\pi}{a\ns_0\sqrt{3}}\,\yhat$.
The three nearest neighbor separation vectors ${\mib\delta}\ns_{1,2,3}$ are given by ${\mib\delta}\ns_1 = -\frac{1}{3}\,\Ba\ns_1 - \frac{1}{3}\Ba\ns_2$, ${\mib\delta}\ns_2 = \frac{2}{3}\,\Ba\ns_1 - \frac{1}{3}\Ba\ns_2$, and ${\mib\delta}\ns_3 = -\frac{1}{3}\,\Ba\ns_1 + \frac{2}{3}\Ba\ns_2$, each of length $|{\mib\delta}\ns_j|= a\ns_0/\sqrt{3} = 1.42\,$\AA.
The in-plane locations of the four sublattices are then given by the subscripts: $u\ns_{\mib R}$, $v\ns_{{\mib R} + {\mib\delta}\ns_1}$, $\util\ns_{{\mib R} + {\mib\delta}\ns_1}$, and $\vtil\ns_{{\mib R} - {\mib\delta}\ns_1}$, and the separation between the $(u, v)$ and $(\util, \vtil)$ planes is $d = 3.35\,$\AA.
The $(\util, \vtil)$ layer (B) is shifted by ${\mib\delta}\ns_1$ relative to the $(u, v)$ layer (A), a configuration known as Bernal stacking.

%%%%%%%%%%%%%%%%%%%%%%%%%%%%%%%%%%%%%%%%%%%%%%%%%%%%%%%%%
%%%%%%%%%%%%%%%%%%%%%%%%%%%%%%%%%%%%%%%%%%%%%%%%%%%%%%%%%

%\begin{table}[!b]
%\begin{center}
%\begin{tabular}{|c||c|c|c|c|c|}\hline
%parameter & $\gamma\ns_0$ & $\gamma\ns_1$ & $\gamma\ns_3$ & $\gamma\ns_4$  & $\Delta'$\\ \hline
%value (eV) & $3.0$ & $0.41$ & $0.3$ & $0.15$ & $0.018$ \\ \hline
%\end{tabular}
%\caption{\label{partab} SWMc parameters (from Ref.~\onlinecite{ZhangL2008dot}).}
%\end{center}
%\end{table}

Note that repeating the Bernal stacking ABABAB$\ldots$ generates the common form of graphite.
In graphite, the $v$ and $\util$ sublattices form one-dimensional chains, while the $u$ and $\vtil$ sites lie above and below hexagon centers in neighboring planes.
The electronic structure of graphite dates to the seminal work of Wallace~\cite{Wallace1947tbt} and subsequent work by McClure~\cite{McClure1957bso} and by Slonczewski and Weiss,~\cite{Slonczewski1958bso} known as the Slonczewski-Weiss-McClure (SWMc) model. The SWMc model is equivalent to a seven-parameter tight binding model which describes nearest neighbor in-plane hopping (amplitude $-\gamma\ns_0$), three interplane hopping processes $(\gamma\ns_1,\gamma\ns_3,\gamma\ns_4$), two next-nearest plane hoppings ($\gamma\ns_2,\gamma\ns_5$), and an on-site energy shift $\Delta'$ which distinguishes the chain sites $(v,\util)$ from the non-chain sites $(u,\vtil)$ in each unit cell. Parameter $\Delta'$ should not be confused with $\Delta \equiv \Delta' + \gamma_2 - \gamma_5$.

In BLG, $\gamma_2$ and $\gamma_5$ do not enter and one further expects~\cite{ZhangL2008dot} $\Delta'_{\rm BLG} = \Delta'_{\rm graphite}\, /\, 2$. Therefore, in BLG we are left with five parameters: $\gamma\ns_{0} = 3.0\,\text{eV}$, $\gamma\ns_{1} = 0.41\,\text{eV}$, $\gamma\ns_{3} = 0.3\,\text{eV}$, $\gamma\ns_{4} = 0.15\,\text{eV}$, and $\Dpr = 0.018\,\text{eV}$. (For the interpretation of these parameters within the tight-binding picture, see Fig.~\ref{fig:bilayer}.
For a discussion of their numerical values, including the uncertainties, see Ref.~\onlinecite{ZhangL2008dot}.) Finally, to describe a biased BLG, we include a scalar potential $\pm U$ on the two layers.

The SWMc Hamiltonian of BLG in second-quantized notation is written as ${\widehat H} = \sum_\Bq\Psi\yd_\Bq\,H\ns_\Bq\,\Psi\ns_\Bq$, where
\begin{equation}
  \Psi\yd_\Bq =
  \begin{pmatrix}
    u\yd_\Bq & v\yd_\Bq & \util\yd_\Bq & \vtil\yd_\Bq
  \end{pmatrix}\,,
\end{equation}
is a four (sublattice) component creation operator with crystal momentum $\Bq$, and
\begin{equation}
  H\ns_\Bq =
  \begin{pmatrix} -U & -\gze\,\Sq & \gfo\,\Sq & \gth\,\Sqs \\
    -\gze\,\Sqs & - U  + \Delta'  & \gon & \gfo\,\Sq \\
    \gfo\,\Sqs & \gon & U + \Delta'  & -\gze\,\Sq \\
    \gth\,\Sq & \gfo\,\Sqs & -\gze\,\Sqs & U
  \end{pmatrix}\,.
  \label{Hbilayer}
\end{equation}
Here, as in Ref.~\onlinecite{Wallace1947tbt}, we define the dimensionless in-plane hopping amplitude
\begin{equation}
\Sq = e^{i\Bq\cdot{\mib\delta}\ns_1} + e^{i\Bq\cdot{\mib\delta}\ns_2} + e^{i\Bq\cdot{\mib\delta}\ns_3} \,.
\label{eqn:wallace}
\end{equation}
In the vicinity of the two inequivalent Brillouin zone corners $\Bq = \pm\BK$ (see Fig.~\ref{fig:monolayer}), $S\ns_\Bq$ vanishes, and writing $\Bq = \pm\BK + \Bk$ one finds
\begin{align}
S\ns_{\BK + \Bk}& = -\frac{\sqrt{3}}{2}\,\big(k\ns_x - i k\ns_y\big)a_0 + \mathcal{O}(k^2)\,,\\
S\ns_{ - \BK + \Bk}& = +\frac{\sqrt{3}}{2}\,\big(k\ns_x + i k\ns_y\big)a_0 + \mathcal{O}(k^2)\,. 
\end{align}
Setting all parameters but $\gamma\ns_0$ to zero, one obtains the monolayer dispersion,
\begin{equation}
\ve\ns_k = \gamma\ns_0 |S\ns_\Bq| = \hbar v\ns_0 |\Bk| + {\cal O}(k^2)\ \,,
\end{equation}
where $v\ns_0 = \sqrt{3} \gze a\ns_0/2\hbar\approx 1.0\times10^8\,{\rm cm}/{\rm s}$ is the Fermi velocity.

If we turn on the interlayer hopping $\gamma\ns_1$ and the interlayer potential $U$, keeping $\gth = \gfo = \Dpr = 0$, then we obtain\footnote{The main effect of finite $\gamma_4$ and $\Dpr$ is to produce a small but measurable electron-hole asymmetry.~\cite{ZhangL2008dot}} the spectrum
\begin{align}
E_{s\ns_1 s\ns_2, k} &= s\ns_1\sqrt{\frac{1}{2}\gamma_1^2 + U^2 + \ve^2  +s\ns_2 \,\sLambda^2(\ve\ns_k)} \,,
\label{eqn:BLG_E_k}\\
\sLambda(\ve)&\equiv \Big[\frac{1}{4}\gamma_1^4 + (\gamma_1^2 + 4 U^2) \, \ve^2 \Big]^{1/4} \,.
\label{eqn:Lambda}
\end{align}
Here $s\ns_1$ and $s\ns_2$ label the four bands as follows: $s\ns_1 = \pm$ labels the conduction and valence bands, respectively, while $s\ns_2 = + 1$ for the outer bands and $s\ns_2 = -1$ for the inner bands.
Thus, the ordering of the four levels is
\begin{equation}
E\ns_{-+} < E\ns_{--} \le E\ns_{+-} < E\ns_{++} \,.
\end{equation}
(For aesthetic reasons, we will usually abbreviate $s\ns_{1,2} = \pm$ in the subscripts, as above.)

Due to particle-hole symmetry at $\gfo = \Dpr = 0$, we may restrict our attention to the conduction bands $s\ns_1 = +1$.
In this case, the shape of the energy bands is as follows.
For the outer band, $E\ns_{++}$ is a monotonic function of $\ve$, starting at $\Bq = \pm\BK$, where $E\ns_{++} = E\ns_\diamond\equiv\sqrt{\gamma_1^2+U^2}$ and extending to $E\ns_{++}(0)\approx 3\gze$ (assuming $\gze\gg\gon,U$). 
We will be interested mainly in the inner ($s\ns_2 = -1$) bands, shaped as the Mexican hats near $\Bq = \BK^\pm$ , i.e., $\Bk = 0$.
For example, the conduction band $E\ns_{+-, k}$ has a local maximum --- the top of the hat --- at $\ve\ns_k = 0$, where $E\ns_{+-} = U$ and a local minimum  --- the bottom of the hat --- at $\ve\ns_k = \ve\ns_\star$, where
\begin{equation}
\vestar = \sqrt{U^2 + E_\star^2} \,, \quad \Estar
 = \dfrac{\gamma\ns_1 U}
        {\sqrt{\gamma_1^2 + 4 U^2}} \,.
\label{eqn:E_*}
\end{equation}
Hence, this minimum is attained on circles of radius $\kstar = \ve\ns_\star/\hbar v\ns_0$, centered at the zone corners.

Inverting the relation between $E$ and $\ve$, and suppressing the labels $s\ns_{1,2}$, one finds
\begin{align}
\ve_k^2& = E_k^2+U^2 - s\ns_3 \sGamma^2(E\ns_k) \,,\\
\sGamma(E)&\equiv \Big[(\gamma_1^2+4U^2)E^2-\gamma_1^2\, U^2\Big]^{1/4},  \label{eqn:Gamma}
\end{align}
where $s\ns_3 = \pm 1$.
This equation has no solutions when $E^2 < E_\star^2$, which is the band gap.
There are two solutions when $\Estar \leq |E| \leq U$, both in the inner ($s\ns_2 = -1$) band.
For $U \leq |E| \leq E\ns_\diamond \equiv \sqrt{U^2 + \gamma_1^2}$, the energy is between the local maximum of the inner band and the minimum of the outer band, and there is one solution.
Finally, for $|E| > E\ns_\diamond$ there are again two solutions, one with $s\ns_2 = -1$ and one with $s\ns_2 = +1$.
As we shall see in Sec.~\ref{Sec:LL_crossing}, the existence of two solutions $E$ within the inner band --- one on the inside and the other on the outside of the Mexican hat --- gives rise to multiple level crossings when magnetic field is turned on.

If one is interested only in the inner bands and low energies, $|E_{s\ns_1, -, k}| \ll \gamma\ns_1$, then one can implement a unitary transformation, discussed in Appendix~\ref{sec:low_energy_H_Eff}, which decouples the inner ($s\ns_2 = -1$) and outer ($s\ns_2 = +1$) bands.~\cite{McCann2006lld, Nilsson2008epo} The results of this procedure are further described in Sec.~\ref{sec:LEET}.

\subsection{Quantizing magnetic field}

In the presence of a magnetic field ${\mib B} = B\zhat$, the two components of the wavevector no longer commute.
Near the zone corners, we invoke the Kohn-Luttinger substitution $k_i \to \pi_i$ such that $[\pi\ns_x, \pi\ns_y] = -i / \ell_{\!B}^2$, where $\elB = \sqrt{\hbar c / e|B|}$ is the magnetic length. We define the ladder operators,
\begin{equation}
  a = -\frac{\elB}{\sqrt{2}}\, (\pi_x - i \pi_y)\,,
  \quad
  a\yd = -\frac{\elB}{\sqrt{2}}\, (\pi_x + i \pi_y)\,,
  \label{eqn:a}
\end{equation}
which satisfy the commutation relation
\begin{equation}
  \big[ a \,, \, a\yd \big] = \sign(B)\,.
  \label{eqn:a_comm}
\end{equation}
Using Eqs.~\eqref{Hbilayer} and \eqref{eqn:a}, we find the Hamiltonian of $\Kp$ valley to be
\begin{equation}
\CH^+(U) = \begin{pmatrix} -U & -\omo\,a & \eta\ns_4\,\omo\,a & \eta\ns_3\,\omo\,a\yd \\
-\omo\,a\yd & -U+\Dpr  & \gon & \eta\ns_4\,\omo\,a \\
\eta\ns_4\,\omo\,a\yd & \gon & U+ \Dpr  & -\omo\,a \\
\eta\ns_3\,\omo\,a & \eta\ns_4\,\omo\,a\yd & -\omo\,a\yd & U \end{pmatrix},
\label{HpB}
\end{equation}
where $\eta\ns_3 = \gth/\gze = 0.1$, $\eta\ns_4 = \gfo/\gze = 0.05$, and
\begin{equation}
\omo = \sqrt{2}\, \dfrac{\hbar v\ns_0}{\elB} \approx 35\,\text{meV} \sqrt{|B(\text{T})|}\,.
\end{equation}
Throughout we shall ignore the effects of real Zeeman splitting, which are small due to the value of the Bohr magneton, $\mu\ns_{\rm B} = e\hbar/2m_{\rm e}c = 57.9\,\mu{\rm eV}/{\rm T}$.
At the highest fields in the relevant experiments ($B\approx 30\,$T) the real Zeeman splitting is on the order of a few millivolts, which is much smaller than even the smallest of the SWMc energy scales. (As we shall see, the pseudo-Zeeman effect can be significantly larger.)

%%%%%%%%%%%%%%%%%%%%%%%%%%%%%%%%%%%%%%%%%%%%%%%%%%%%%%%%%
\begin{figure}[t]
  \begin{center}
    \includegraphics[width=3.2in]{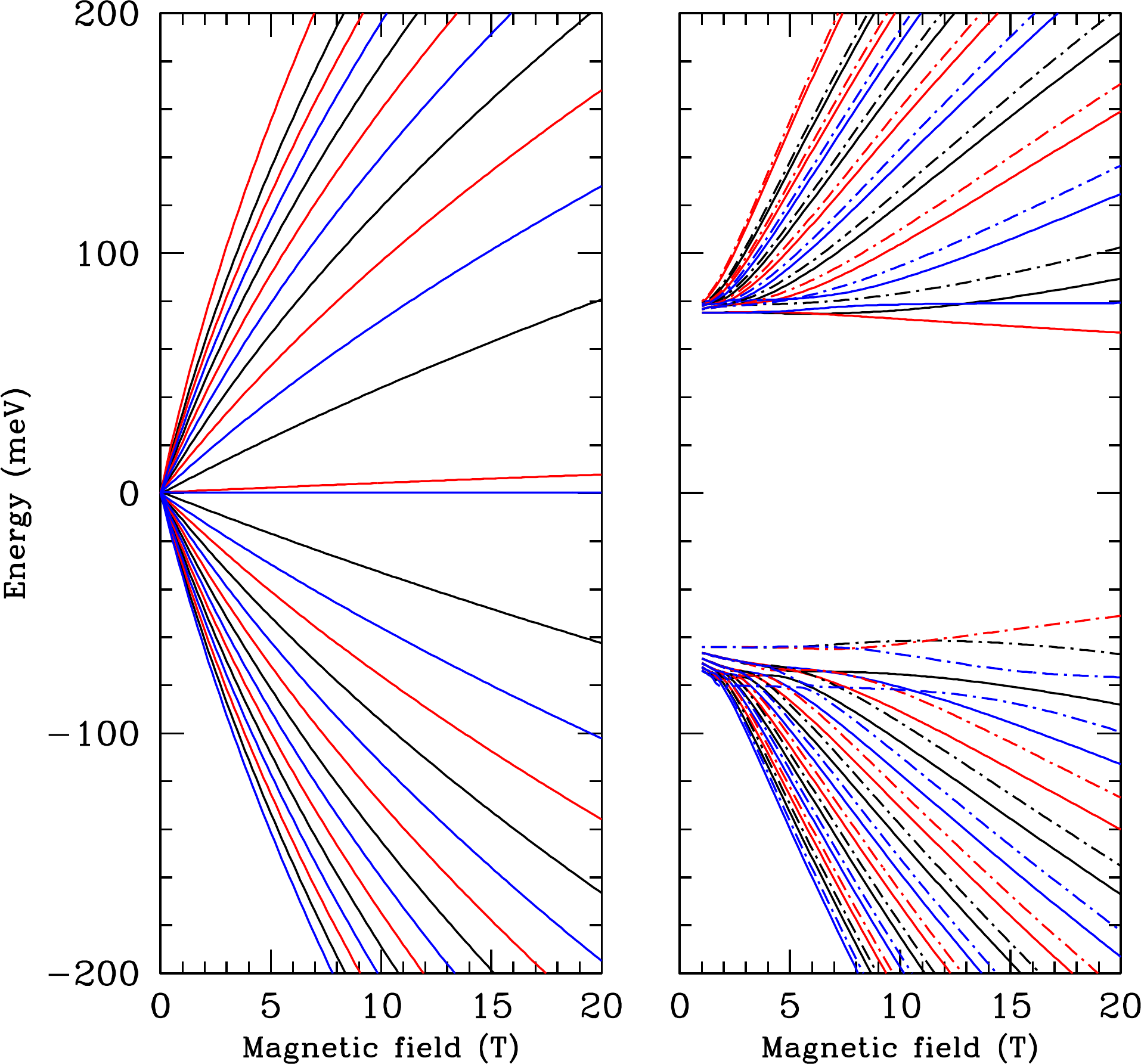}
  \end{center}
  \caption{(Color online) Landau level energies {\it vs.} magnetic field for $U = 0$ (left) and $U = 80\,$meV (right).
  Solid lines correspond to the $K^+$ valley and broken lines to the $K^-$ valley.
  The color distinguishes the spectra of $\CH_{\rm a}$ (black), $\CH_{\rm b}$ (red), and $\CH_{\rm c}$ (blue), where $\CH_{\rm a,b,c}$ are defined in Appendix B.}
  \label{fig:hspec}
\end{figure}
%%%%%%%%%%%%%%%%%%%%%%%%%%%%%%%%%%%%%%%%%%%%%%%%%%%%%%%%%

The Hamiltonian $\CH^-$ of $\Km$ valley is obtained from $\CH^+$ via the replacements
\begin{equation}
  R_x:\quad a \to -a\yd\,, \quad a\yd \to -a\,,
  \label{eqn:a_repl}
\end{equation}
which is the reflection in the $y$--$z$ plane.
The commutation relation~\eqref{eqn:a_comm} between $a$ and $a\yd$ and therefore the energy spectrum is preserved if we additionally reverse the magnetic field,
\begin{equation}
  R_B:\quad B \to -B\,.
\label{eqn:B_reverse}
\end{equation}
Taken together, these replacements implement the symmetry operation $O_1 = R_x R_B$ discussed in Sec.~\ref{sec:Introduction}.
The other valley-interchanging operator $O\ns_2$ is represented in terms of the unitary matrix
\begin{equation}
\mathcal{V} = \begin{pmatrix} 0 & \sigma_x \\ \sigma_x & 0 \end{pmatrix}\,
\end{equation}
and the time-reversal operation $S_\Bq \to (S_\Bq)^*$, i.e.,
\begin{equation}
R_T:\quad a \to -a\,, \quad a\yd \to -a\yd\,, \quad B \to -B\,.
\label{eqn:time_reversal}
\end{equation}
It is easy to see that
\begin{equation}
\CHm(U) = R_T R_B \left[ \mathcal{V}\yd\, \CHp(-U) \,
 \mathcal{V} \right].
\end{equation}
Since $R_T R_B$ also does not change the commutation relation~\eqref{eqn:a_comm}, the spectra of $\CHm(U)$ and $\CHp(-U)$ coincide.
Thus, it suffices to discuss the spectrum of $\CHp$, from which one can obtain the spectrum of $\CHm$ by reversing the sign of either $U$ or $B$.

These symmetries further imply that at $B = 0$ the two valleys are degenerate in energy and that additionally, each valley is symmetric under $U\to -U$.
On the other hand, at finite $B$, the valleys are degenerate only if $U = 0$.
Note also that the total spectrum, including both valleys, is particle-hole symmetric when $\Dpr = \gfo = 0$.

\begin{figure}[b]
  \begin{center}
    \includegraphics[width=3.2in]{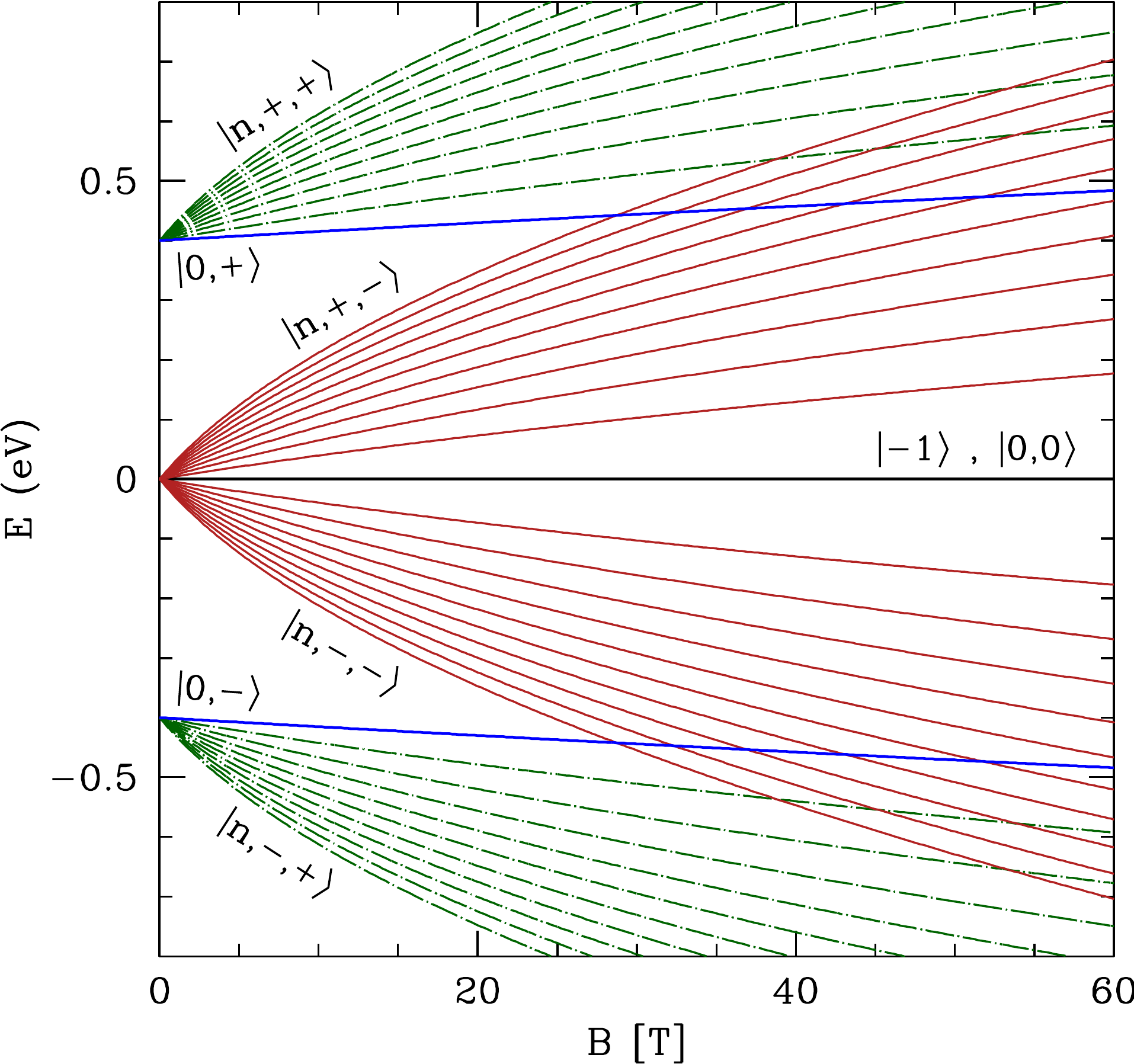}
  \end{center}
  \caption{(Color online) Landau level energies {\it vs.} magnetic field $B$ for the case $\gamma_3 = \gamma_4 = U = 0$.
  The labeling of the states corresponds to that in the text.}
  \label{fig:states}
\end{figure}

Making use of the eigenvectors $\ket{n}$ of the number operator $a\yd a$, we write the general bilayer wavefunction as
\begin{equation}
  \ket{\Psi} = \sum_{n = 0}^\infty
  \begin{pmatrix}
    u\ns_n \ket{n} \\ v\ns_n \ket{n} \\ \util\ns_n \ket{n} \\ \vtil\ns_n \ket{n}
  \end{pmatrix}\,.
\end{equation}
The matrix representation of the corresponding Hamiltonian is discussed in Appendix B. If all SWMc parameters are kept, it can be diagonalized only numerically. Some results are shown in Figs.~\ref{fig:hspec}, \ref{fig:states}, and \ref{fig:vspecb}, which illustrate that the spectrum can be rather complicated. In the remainder of this section we review certain limits where some analytical progress can also be made, which helps with physical understanding of these results.

%%%%%%%%%%%%%%%%%%%%%%%%%%%%%%%%%%%%%%%%%%%%%%%%%%%%%%%%%
%%%%%%%%%%%%%%%%%%%%%%%%%%%%%%%%%%%%%%%%%%%%%%%%%%%%%%%%%

\subsubsection{$\gamma_3 = 0$ limit}

It is simplest to consider the case where $\gth = 0$, which turns out to be an excellent approximation at large fields.
When $\gth = 0$, the eigenstates of $\CHp$ fall into one of three classes:
\begin{equation}
  \ket{\psi\ns_{-1}}=
  \begin{pmatrix}
    0 \\ 0 \\ 0 \\ \ket{0}
  \end{pmatrix}\quad,\quad
  \ket{\psi\ns_0}=
  \begin{pmatrix}
    0 \\ v\ns_0\ket{0} \\ \util\ns_0 \ket{0} \\ \vtil\ns_1\ket{1}
  \end{pmatrix}\,,
\end{equation}
and
\begin{equation}
  \ket{\psi\ns_n} =
  \begin{pmatrix}
    u\ns_{n-1}\ket{n-1} \\ v\ns_n\ket{n} \\ \util\ns_n \ket{n} \\ \vtil\ns_{n+1}\ket{n+1}
  \end{pmatrix}\,,
\end{equation}
with $n\ge 1$.
Clearly $\ket{\psi\ns_{-1}}$ is an eigenstate with eigenvalue $E = U$.
Applying $\CHp$ to $\ket{\psi\ns_0}$, one obtains the $3\times 3$ Hamiltonian for the $\psi\ns_0$ sector,
\begin{equation}
H\ns_0 = \begin{pmatrix}-U+\Dpr & \gon & \efo\,\omo \\ \gon & U + \Dpr & -\omo \\
\efo\,\omo & -\omo & U \end{pmatrix}\,.
\end{equation}
Finally, the spectrum for the $\ket{\psi\ns_n}$ states ($n\ge 1$) is given by the eigenvalues of the $4\times 4$ Hamiltonian
\begin{equation}
  H\ns_n =
  \begin{pmatrix}
    -U & -W\ns_n & \efo W\ns_n & 0 \\
    -W\ns_n & -U + \Dpr  & \gon & \efo W\ns_{n+1} \\
    \efo W\ns_n & \gon & U + \Dpr  & -W\ns_{n+1}\\
    0 & \efo W\ns_{n+1}  & -W\ns_{n+1}\ & U
  \end{pmatrix}\,,
  \label{Hnblock}
\end{equation}
where
  \begin{equation}
    W\ns_n \equiv \sqrt{n}\>\omega\ns_0\,.
  \end{equation}
%%

%%%%%%%%%%%%%%%%%%%%%%%%%%%%%%%%%%%%%%%%%%%%%%%%%%%%%%%%%
\begin{figure}[t]
\begin{center}
\includegraphics[width=3.2in]{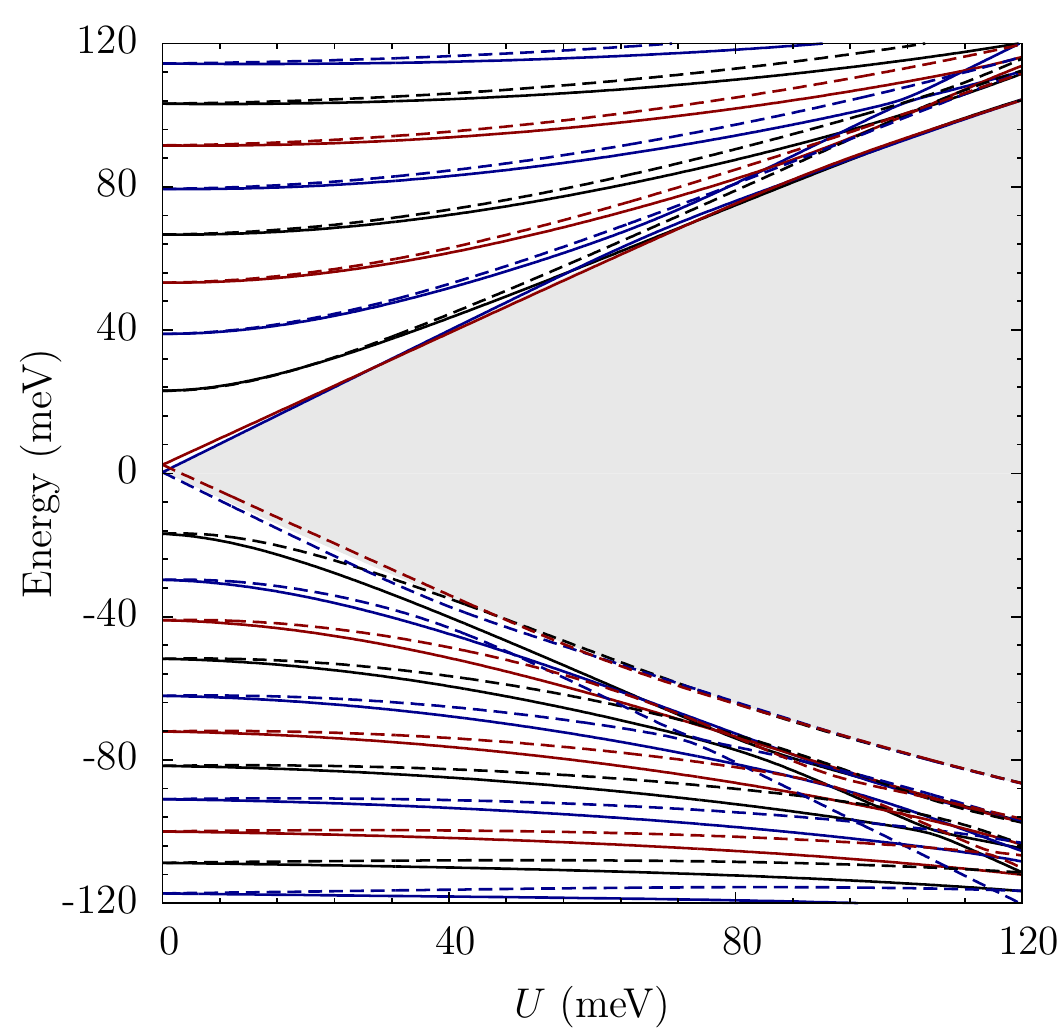}
\end{center}
\caption{(Color online) Landau level energies {\it vs.} interlayer bias $U$ for a field value $B = 5\,$T.
Solid lines correspond to the $K^+$ valley and broken lines to the $K^-$ valley.
The color distinguishes the spectra of $\CH_{\rm a}$ (black), $\CH_{\rm b}$ (red), and $\CH_{\rm c}$ (blue).
At an accidental degeneracy (level crossing), the color and the line type cannot both be identical.
The shaded area indicates the energy gap at $B = 0$.}
\label{fig:vspecb}
\end{figure}
%%%%%%%%%%%%%%%%%%%%%%%%%%%%%%%%%%%%%%%%%%%%%%%%%%%%%%%%%

To label the states, it is helpful to consider the case $\gth = \gfo = \Dpr = U = 0$, corresponding to a pure nearest-neighbor hopping model with constant (zero) local site energies.
One then finds the following (valley-degenerate) spectrum:
\begin{align}
&\CE\ns_{-1} = \CE\ns_{0} = 0\,,\quad
\CE\ns_{0,s\ns_1,+} = s\ns_1\gon\sqrt{1+\beta\,} \,,\\
&\CE\ns_{n,s\ns_1,s\ns_2} = s\ns_1\gon\sqrt{\half+\left(n+\half \right) \, \beta + s\ns_2 C\ns_n\,}\,,
\label{purespec}
\end{align}
where
\begin{equation}
\beta = \left(\dfrac{\omo}{\gon}\right)^{\!\!2} = \dfrac{B}{136\,\text{T}}
\label{eqn:beta}
\end{equation}
and
\begin{equation}
C\ns_n = \sqrt{\dfrac{1}{4} + \left(n + \half\right) \,\beta + \dfrac{1}{4} \beta^2}\,.
\end{equation}
%%
% A valley label ($\pm$) is then added to provide an indexing of the
% complete spectrum.
Note that the $n = -1$ and $n = 0$ states require a separate labeling convention.

For the full model, with $\gth$, $\gfo$, and $\Dpr$ restored, particle-hole symmetry is broken by the $\gfo$ and $\Dpr$ terms.
These are relatively small however, so there remains an approximate particle-hole symmetry, as shown in Fig.~\ref{fig:vspecb}.
The state labels are then defined by adiabatic continuity with the $\gth = \gfo = \Dpr = 0$ limit.

\subsubsection{Low energy effective theory}
\label{sec:LEET}

As mentioned above, at low energies one can implement a unitary transformation, which decouples the inner ($s\ns_2 = -1$) and outer ($s\ns_2 = +1$) bands order by order in $S$, which vanishes at the zone corners.~\cite{McCann2006lld, Nilsson2008epo} Here $S = \mp (\sqrt{3} / 2) a_0 (k_x \mp i k_y)$ and the upper (lower) sign denotes the $K^+$ ($K^-$) valley.
To order $S^2$ one obtains
\begin{equation}
\Htil =
\begin{pmatrix} \lambda \big(\Dtil + 2U\big)\,S S\yd - U & \gth S\yd - \dfrac{\gamma_0^2}{\gon}\,S^2 \\ & \\
\gth S - \dfrac{\gamma_0^2}{\gon} \,S\yd{}^2 & \lambda \big(\Dtil - 2U\big)\,S\yd S + U
\end{pmatrix},
\label{eqn:H_eff_full}
\end{equation}
where $\lambda = (\gze/\gon)^2 \approx 53.5$ and
\begin{equation}
\Dtil = \Dpr + \dfrac{2\gon\gfo}{\gze} \approx 59\,{\rm meV}
\label{eqn:Delta_tilde}
\end{equation}
is a composite parameter describing electron-hole symmetry breaking effects of $\Dpr$ and $\gfo$. Anticipating the introduction of an external magnetic field, we have allowed for the possibility that $S$ and $S\yd$ do not commute, cf.~Appendix~\ref{sec:low_energy_H_Eff} for derivation.

The eigenvalues of $\Htil$ to leading order in $\gth$ are~\footnote{This is similar to Eq.~(16) of Ref.~\onlinecite{Nilsson2008epo} except in lieu of our $\cos 3 \varphi$ they have $-\cos 3 \phi = \sin 3 \varphi$.} 
\begin{equation}
\begin{split}
\Etil_{s\ns_1, -,\Bk}^{\pm} &= \dfrac{\ve_k^2}{\gamma_1^2} \, \Dtil\\
&+ s\ns_1\sqrt{\left( \dfrac{2\ve_k^2}{\gamma_1^2} -1 \right)^{\!\!2} U^2 + 
\dfrac{\ve_k^4}{\gamma_1^2} \pm \dfrac{2\gth\,\ve_k^3}{\gze\gon} \cos 3\vphi} \,,
\end{split}
\label{eqn:b_square}
\end{equation}
where $\varphi$ is the polar angle of $\Bk$.
This agrees with Eq.~\eqref{eqn:BLG_E_k} in the appropriate limit.

The $2 \times 2$ form of matrix $\Htil$ in Eq.~\eqref{eqn:H_eff_full} naturally leads to the concept of a pseudospin-$\half$ degree of freedom which simplifies calculations somewhat.
We use this approach sparingly for the following reasons. 
First, in experiments $U$ is not necessarily much smaller than $\gon$, in which case the reduction to a two-band effective Hamiltonian is not valid.
Second, the calculation of the pseudospin-related effects are not difficult even when all four bands are kept.
Finally, the low-energy theory does not produce accurate results for the Berry phase.  
A brief discussion of this technical issue is also given in Appendix~\ref{sec:low_energy_H_Eff}.   

\begin{figure}[b]
  \centering
  \includegraphics[width = 3.2in]{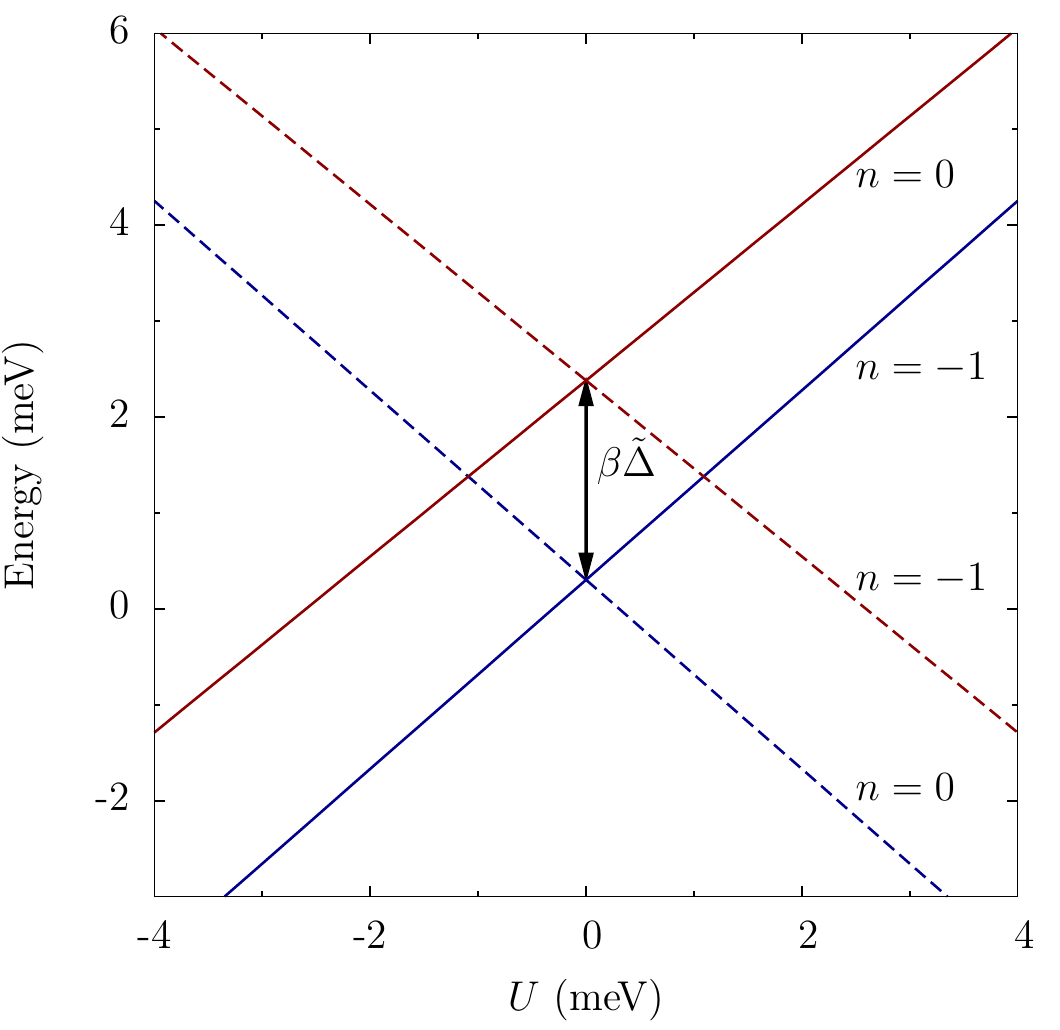}
  \caption{(Color online) Landau levels $-1$ and $0$ as a function of $U$ at $B = 5\,\text{T}$, i.e., $\beta = 0.037$.
The solid (dashed) lines correspond to $K^+$ ($K^-$) valley.
  The SWMc parameters are taken from Ref.~\onlinecite{ZhangL2008dot} and
  $\Dtil = 59\,\text{meV}$, cf.~Eq.~\eqref{eqn:Delta_tilde}.
  \label{fig:BLG_LL_Vb_zoom}}
\end{figure}

In a nonzero magnetic field, $\Htil^+$ becomes
\begin{equation}
\begin{pmatrix} \beta\big(\Dtil + 2U\big)\,a a\yd - U & \dfrac{\gth}{\gze}\,\omo\, a\yd - \beta\gon a^2 \\ & \\
\dfrac{\gth}{\gze}\,\omo\, a - \beta\gon a\yd{}^2 & \beta\big(\Dtil - 2U\big)\,a\yd a + U \end{pmatrix},
\label{eqn:HeffB}
\end{equation}
while $\Htil^-$ is obtained via substitutions~\eqref{eqn:a_repl}.
When $\gth = 0$, their eigenvalues are easily obtained by considering the basis of states
\begin{equation}
\ket{\phi^+_n} = \begin{pmatrix} u^+_n \ket{n-1} \\ v^+_n  \ket{n+1} \end{pmatrix}\,.
\end{equation}
In this basis the above Hamiltonian takes the form
\begin{equation}
\begin{pmatrix} \beta\, (\Dtil + 2U) \, n - U & -\beta\gon\sqrt{n(n+1)} \\ & \\
-\beta\gon\sqrt{n(n+1)} & \beta\, (\Dtil - 2U) (n+1) + U \end{pmatrix}.
\label{Htilp}
\end{equation}
When $n = -1$, we have $u^+_{-1} = 0$ and the energy levels in the two valleys are $\CE_{-1}^\pm = \pm U$.
With $n = 0$ we again have $u^+_0 = 0$, and
\begin{equation}
\CEtil^\pm_{0} = \beta \Dtil \pm (1 + 2\beta)U\,.
\label{eqn:E_0}
\end{equation}
The splitting of the $n = -1$ and $n = 0$ levels and their valley-dependent slope as a function of $U$ lead to a characteristic diamond-shaped crossing pattern, shown in Fig.~\ref{fig:BLG_LL_Vb_zoom}.
The largest energy gap occurs in the unbiased sample, $U = 0$, and its magnitude $\approx 0.5\,\text{meV} \times B(\text{T})$ is comparable to that measured in Ref.~\onlinecite{Feldman2009bss} in a suspended BLG.
On the other hand, an order of magnitude smaller gaps (smaller than even the bare Zeeman gap) have been observed in a more disordered BLG on SiO$_2$ substrate.~\cite{Zhao2010sbi}

Finally, for $n > 0$ one has (similar to Ref.~\onlinecite{Koshino2010pav})
\begin{equation}
\begin{split}
\CEtil^\pm_{n,s\ns_1,-} & = \bigg(n + \half \bigg) \beta \Dtil \mp \beta U\\
 &+ s\ns_1 \sqrt{ \Big[ (2n + 1)\beta U \mp \dfrac{\beta \Dtil}{2} - U \Big]^2 \!\! + n (n + 1)\,\beta^2 \gamma_1^2\,} \,.
\end{split}
\label{eqn:E_low_energy}
\end{equation}
This completes our summary of the (mostly) known analytic results for the energy spectrum of BLG.

%%

%%%%%%%%%%%%%%%%%%%%%%%%%%%%%%%%%%%%%%%%%%%%%%%%%%%%%%%%%%%%%%%%%%%%%%%%%%%%%%%
\section{Quasiclassical approximation}
\label{sec:quasiclassical}
\subsection{Effective $\boldsymbol{g}$-factor}
\label{sec:g_factor}
%%%%%%%%%%%%%%%%%%%%%%%%%%%%%%%%%%%%%%%%%%%%%%%%%%%%%%%%%%%%%%%%%%%%%%%%%%%%%%%

Renormalization of the electron magnetic moment is a well-known phenomenon in the solid-state physics.  
Most often it comes from spin-orbit interaction; however, in crystals without inversion symmetry there is an additional contribution due to the orbital angular momentum:  
\begin{equation} 
{\mib M}_\alpha \equiv \expect{\alpha} {{\mib M}} {\alpha} = -\dfrac{e}{ 2c}
  \bra{\alpha} {\mib r} \times {\mib v} \ket{\alpha}\,.
  \label{eqn:M_def}
\end{equation}
Here $\alpha$ is a given Bloch state and ${\mib r}$, ${\mib v}$ are the position and velocity operators, respectively.
Since we are not interested in the center-of-mass motion, in evaluating ${\mib M}_\alpha$ we must assume that the expectation value of position vanishes, i.e., that ${\mib r}$ has only off-diagonal matrix elements~\cite{Lifshitz1980sp2}
\begin{equation}
\expect{\alpha} {\mib r} {\alpha'} = i \expect{\alpha}{\Bnab\ns_k}{\alpha^\prime}\,,
\quad \alpha \neq \alpha^\prime\,.
\label{eqn:r_alphabeta}
\end{equation}
This leads to
\begin{equation}
{\mib M}_\alpha = \dfrac{e}{ 2ic} \, {\sum\limits_{\alpha'\neq\alpha}}
\left[\expect{\alpha}{\Bnab\ns_k}{\alpha'} \times \expect{\alpha'}{\mib v}{\alpha}\right]\,.
\label{eqn:M_Chang}
\end{equation}
A lucid derivation of Eq.~\eqref{eqn:M_Chang} was given previously in
Refs.~\onlinecite{Chang1996bph, Sundaram1999wpd}, which also contain references to much earlier work.~\footnote{Unfortunately, some of these sources also contain typographic mistakes.
For example, Eq.~(59.11) of Ref.~\onlinecite{Lifshitz1980sp2} is off by the factor of $m \hbar$ and Eq.~(3.6) of Ref.~\onlinecite{Sundaram1999wpd} is missing a factor of two.}

Below we assume that ${\mib B}$ and ${\mib M}$ are both in the $\zhat$-direction.
The orbital contribution to the $g$-factor is $g = 2 M_\alpha / \mu\ns_{\rm B}$ where $\mu\ns_{\rm B} = e \hbar / (2 m_{\rm e} c)$ is the Bohr magneton and $m_{\rm e}$ is the bare electron mass.
To calculate $M_\alpha$, we can add and subtract the omitted diagonal term in Eq.~\eqref{eqn:M_Chang}, which gives
\begin{equation}
M_\alpha = \dfrac{e}{ 2c} \, (F_\alpha - D_\alpha)\,,
\label{eqn:M}
\end{equation}
where
\begin{align}
F_\alpha &= -i \expect{\alpha} {\Bnab\ns_k \times {\mib v}}{\alpha} \cdot \zhat\,,
\label{eqn:F}\\
D_\alpha &= -i\, [\bra{\alpha} \Bnab\ns_k \ket{\alpha} \times \Bvg] \cdot \zhat\,,
\label{eqn:D}
\end{align}
(note that both $F_\alpha$ and $D_\alpha$ are real) and where
\begin{equation}
\Bvg \equiv \bra{\alpha} {\mib v} \ket{\alpha}= \hbar^{-1} \Bnab\ns_k\, E_\alpha
\label{eqn:v_g_vector}
\end{equation}
is the group velocity vector (the subscript $\alpha$ in $\Bvg$ is omitted for simplicity). Using these formulas we compute the energy dispersion
\begin{equation}
\Etil_\alpha = E_\alpha - BM_\alpha \,.
\label{eqn:E_shifted}
\end{equation}

It is interesting to compare our formula with those in literature.
A very close analogy is provided by Bi, whose effective Hamiltonian is also a $4 \times 4$ matrix linear in ${\mib k}$.
In an early paper~\cite{Falkovsky1966qco} where the calculation of the $g$-factor of Bi is discussed, the subtraction of the diagonal term $D_\alpha$ is lacking, so that the result is not gauge-invariant.
Below we show that $D_\alpha$ is related to the Berry phase, which apparently has not been handled correctly in Ref.~\onlinecite{Falkovsky1966qco} (considering that it precedes Berry's work~\cite{Berry1984qpi} by almost two decades, it is hardly surprising).  

Let us now apply our general formula to BLG.
For $K^+$ valley we can choose the eigenvectors of $H^{+}$ in the form 
\begin{equation}
\ket{\alpha^{+}} = (u_\alpha e^{-i \varphi}, \, v_\alpha \,, \, \util_\alpha \,, \, \vtil_\alpha \, e^{i \varphi})^{\textsf T},
\label{eqn:psi_B_0}
\end{equation}
where $\alpha$ now throughout this section stands for $\{s\ns_1,s\ns_2,k\}$.
It is assumed that the imaginary parts and the entire dependence on $\varphi$ --- the polar angle of $\Bk$ --- enter via the exponential factors only.
A straightforward calculation yields:
\begin{align}
  D_\alpha^{+} &= \vgr  \, \dfrac{u_\alpha^2 - \vtil_\alpha^2 }{ k}
                = -\dfrac{\vgr U}{  2kE_\alpha}
  \Bigg( 1 + \dfrac{4\ve^2 - \gamma_1^2 }{ 2 s\ns_2 \sLambda^2(\ve\ns_k)} \Bigg)\,,
  \label{eqn:D_BLG}\\
  F_\alpha^{+} &= 2\, v\ns_0 \,
  \dfrac{u_\alpha v_\alpha - \util_\alpha \vtil_\alpha}{k} - D_\alpha^{+}
                = -\dfrac{2 \hbar\, v_0^2 \,  U }{s\ns_2 \sLambda^2(\ve\ns_k)} - D_\alpha^{+}\,,
\label{eqn:F_BLG}
\end{align}
where $\sLambda(\varepsilon)$ is given by Eq.~\eqref{eqn:Lambda}.
The eigenvectors for $K^+$ valley can be obtained by replacing $e^{\pm i\varphi}$ in Eq.~\eqref{eqn:psi_B_0} with $-e^{\mp i\varphi}$ and so the signs of $F_\alpha$ and $D_\alpha$ are reversed.

%%
%% FIG. orbital magnetization
\begin{figure}[t]
  \begin{center}
    \includegraphics[width=3.2in]{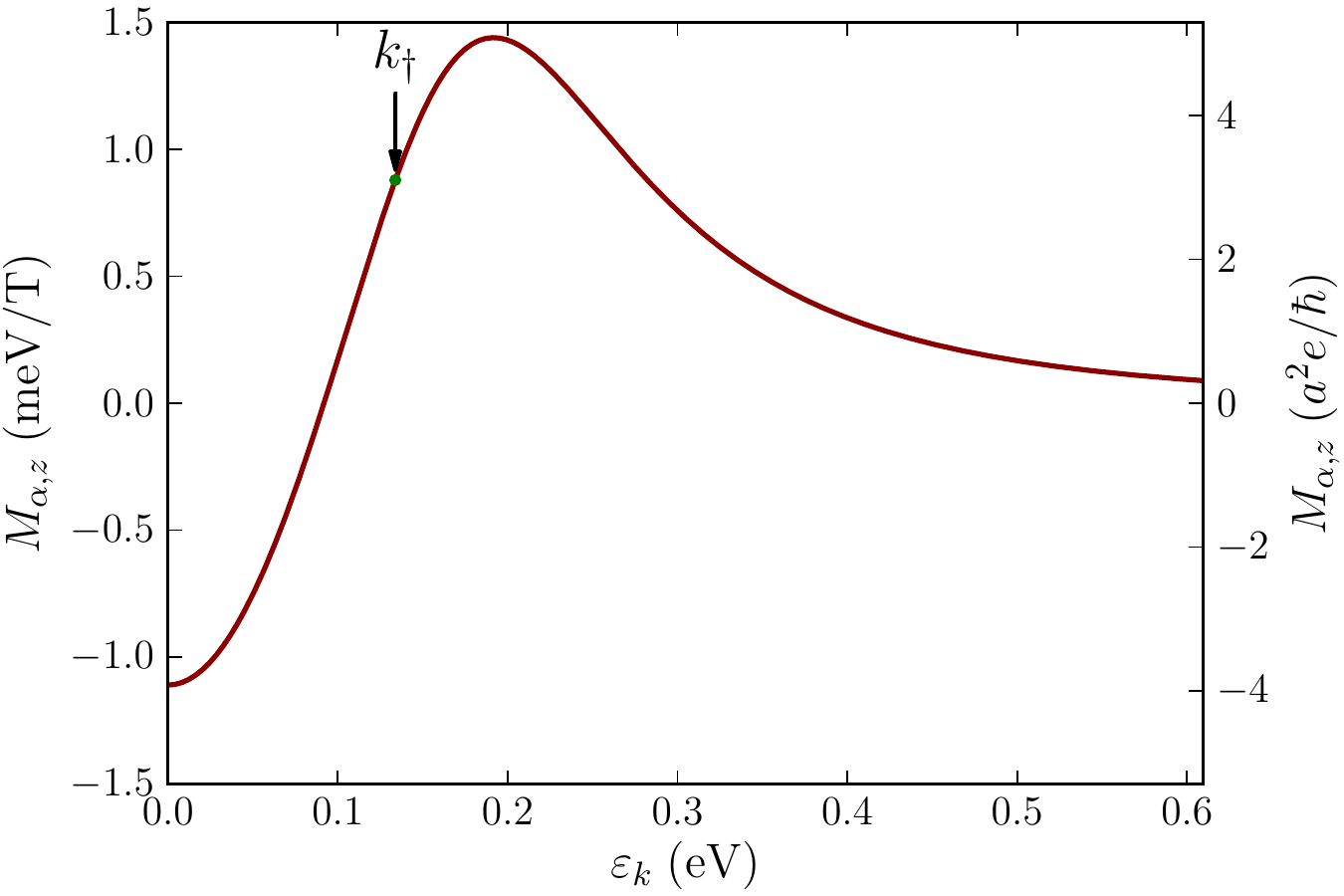}
  \end{center}
  \caption{(Color online) Orbital magnetization $M_{+-}^{+}$ of $K^+$ valley as a function of $\ve_k = \hbar v_0 k$ at $U = 0.1\,\text{eV}$. The location of the band bottom $\ve\protect\ns_k = \ve\protect\ns_\star$ is marked by the arrow.}
  \label{fig:Mz}
\end{figure}
The last term represents the pseudo-Zeeman effect due to the orbital magnetic moment.
Algebraic manipulations with Eqs.~\eqref{eqn:Lambda},
\eqref{eqn:M}, \eqref{eqn:D_BLG}, and \eqref{eqn:F_BLG}, together with the relations
\begin{equation}
  \vgr = \dfrac{1}{\hbar}
  \dfrac{d E}{ dk}
       = v\ns_0 \,\dfrac{\ve}{ E}\,
  \dfrac{s\ns_3 \sGamma^2(E)}{ s\ns_2 \sLambda^2(\ve)}
\label{eqn:v} 
\end{equation}
and
\begin{equation}
  s\ns_3\sGamma^2(E_\alpha) - s\ns_2 \sLambda^2(\ve\ns_k)
  = \half\gamma_1^2 + 2U^2\,,
\label{eqn:GamLam}
\end{equation}
yields
\begin{equation}
M_\alpha^+ = -\dfrac{e \hbar}{c}
\dfrac{2 v_0^2 \gamma_1^2 \, U}{\gamma_1^4 + 4 (\gamma_1^2 + 4 U^2) \, \ve^2_k} \left( 1 - \dfrac{\ve_k^2}{E_\alpha^2} \right) \,.
\label{eqn:orbital_m}
\end{equation}
For the lower energy conduction band, on which we mostly focus later, $M^+_{+-, k}$ is plotted in Fig.~\ref{fig:Mz}.
The modified spectrum $\Etil_\alpha$ is plotted alongside $E_\alpha$ in Fig.~\ref{fig:BLG_E_bands} for all four bands and in Fig.~\ref{fig:BLG_LL_evolution} for the lower conduction band only. At $k = 0$ we have a particularly simple result,
\begin{equation}
g^\pm_{s_1, s_2, 0} = \dfrac{2}{\mu\ns_{\rm B}} \, M^\pm_{s_1, s_2, 0}
= \mp 8 m_\text{e} v_0^2\, \frac{U}{\gamma_1^2}
\label{eqn:M_at_0}
\end{equation}
for all $s\ns_{1, 2}$, in agreement with Eq.~(54) of Ref.~\onlinecite{Koshino2010aom}.

As one can see from Fig.~\ref{fig:Mz}, the $g$-factor has an intriguing energy dependence, which prompts the question of whether it can be verified experimentally. Unfortunately, this appears problematic. There is no optical transition between the energy levels split by the pseudo-Zeeman effect as they belong to different valleys, and so, methods analogous to the electron spin resonance would not work. Another conventional method of extracting the $g$-factor would be to measure the valley-splitting of the Shubnikov-de Haas effect. However, this splitting also includes the contribution of the Berry phase, discussed later in this Section. This contribution effectively compensates for nonmonotonic variation of the $g$-factor, making the valley-splitting of Landau levels only weakly dependent on the Fermi energy (or Landau level index).

The most easily observable manifestation of the pseudo-Zeeman effect appears to be the displacement of the band edges, e.g., the bottom of the Mexican hat of the conduction band. At this point, Eq.~\eqref{eqn:orbital_m} yields (the superscript denotes the valley, as usual):
\begin{equation}
  g^{\pm}_{+-,\kstar} = \dfrac{2}{\mu\ns_{\rm B}} \, M^\pm_{+-,\kstar}
  = \pm \dfrac{8 m_{\rm e} v_0^2 U}{\gamma_1^2 + 4 U^2}\,.
  \label{eqn:g_z_hat}
\end{equation}
Thus, at $U = 100\,\text{meV}$ we obtain $|g^{\pm}_{+-,\kstar}| \approx 22$.
This is one order of magnitude higher than the bare value $g = 2$ and is about as large~\cite{Falkovsky1966qco} as in Bi. (For this reason, we neglect the bare Zeeman coupling in this article.) For $U \ll \gamma\ns_1$, the effective $g$-factor is proportional to $U$, as appropriate for the linear magnetoelectric coupling [Eq.~\eqref{eqn:pseudo-Zeeman}].
Therefore, a roughly linear variation of the band edge positions with $B$ and $U$ can be expected.
This issue is addressed in more detail in Sec.~\ref{sec:LL_spectrum}.

%%%%%%%%%%%%%%%%%%%%%%%%%%%%%%%%%%%%%%%%%%%%%%%%%%%%%%%%%%%%%%%%%%%%%%%%%%%%%%%
\subsection{Quantization rules}
\label{sec:semiclassical_BLG_LL}
%%%%%%%%%%%%%%%%%%%%%%%%%%%%%%%%%%%%%%%%%%%%%%%%%%%%%%%%%%%%%%%%%%%%%%%%%%%%%%%

While numerical calculations of the Landau level spectrum is possible for any choice of parameters, in Sec.~\ref{sec:LL_spectrum} we shall see that the result can be rather complicated.
Therefore, both exact and approximate analytical methods remain valuable for this task in hand.
So far, we have discussed two such methods.
First, for $U = \gfo = \Dpr = 0$, closed-form expressions for the Landau level energies [Eq.~\eqref{eqn:Lambda}] exist.
Second, if these energies are much smaller than $\gon$, then the approximate Eq.~\eqref{eqn:E_low_energy}, valid for finite $U$, can be used.
In this section we outline another approach --- the quasiclassical quantization --- which can be used for arbitrary relation between $U$ and $\gon$.
Within this approximation, Landau level energies $\CE^\pm_{n,s\ns_1,s\ns_2}$ are taken to be equal to the renormalized band energies~\eqref{eqn:E_shifted} evaluated at certain quantized orbits in the reciprocal space:
\begin{equation}
\CE^\pm_{n,s\ns_1,s\ns_2} = E^\pm_{s\ns_1,s\ns_2, k_n^\pm}\,.
\label{eqn:EE}
\end{equation}
If we ignore $\gth$, the orbits are circular and the area of the $n${th} such orbit satisfies the Onsager condition~\cite{Onsager1952iot}
\begin{equation}
  \pi (k_n^\pm \ellB)^2 = 2 \pi (n + \delta_n^\pm)\,,
  \label{eqn:rule_I}
\end{equation}
where $k_n^\pm$ is the radius of the orbit and $\delta_n^\pm$ is a dimensionless number discussed below.

The quasiclassical approximation is accurate through the order ${\cal O}(\ell_{\!B}^{-2})$ or alternatively ${\cal O}(1 / n)$.
It turns out to be exact for parabolic dispersion (where $\delta = \half$) and in monolayer graphene (where $\delta = 0$).
The quasiclassical approximation for general matrix Hamiltonians was previously studied in Refs.~\onlinecite{Littlejohn1991gpa, Littlejohn1991gpi} and specifically in the context of graphene in Ref.~\onlinecite{Carmier2008bpi}.
However, we found it most instructive to follow Refs.~\onlinecite{Chang1996bph, Gosselin2008sqo}.

\begin{figure}[b]
  \centering
  \includegraphics[width=3.3in]{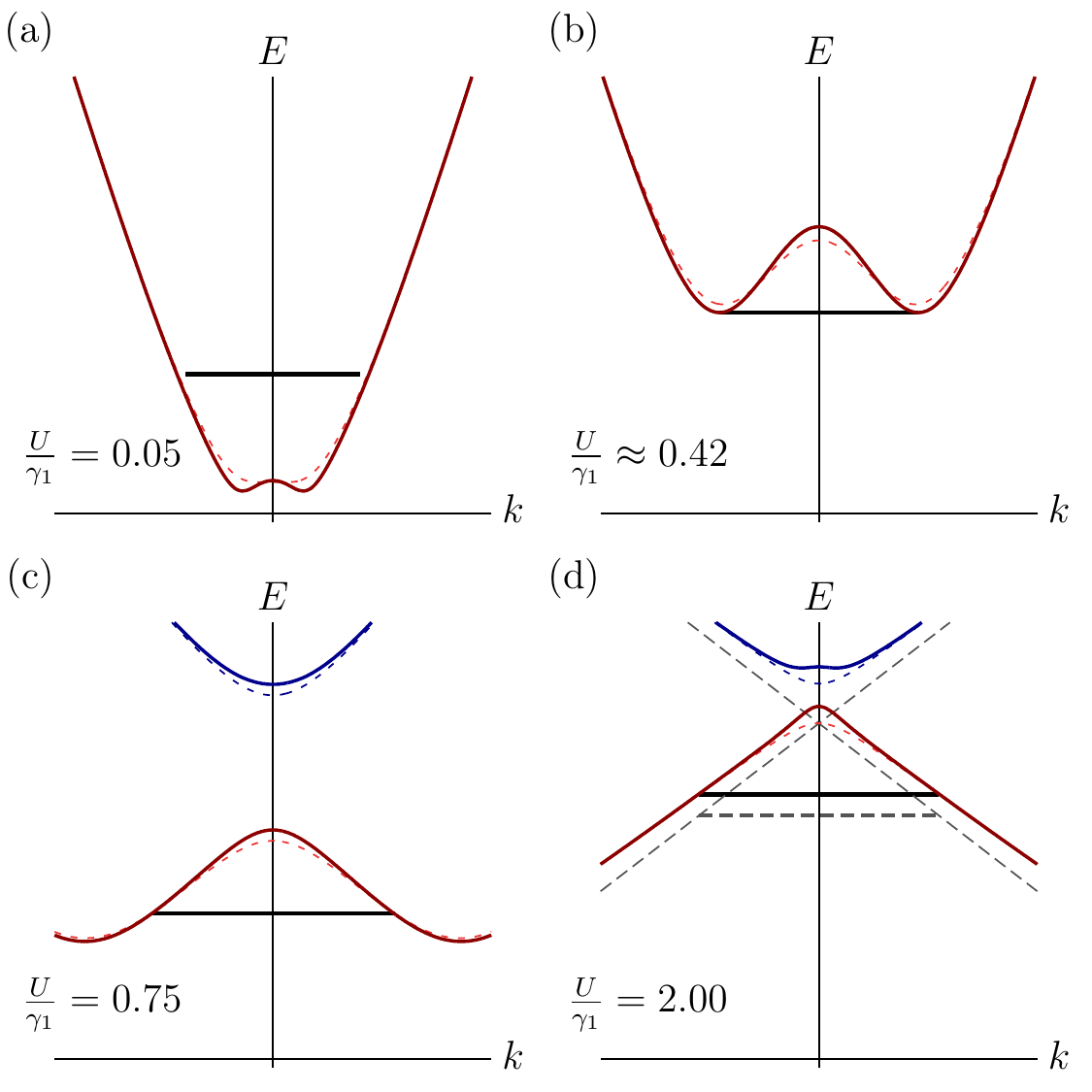}
  \caption{(Color online) Evolution of a particular ($n = 5$) Landau level of the $K^+$ valley as a function of $U$.
  Superimposed are the spectra at zero field (thin traces) and that with pseudo-Zeeman correction in a magnetic field $B = 5\,\text{T}$ (thick trace).
  (a) At small $U$, the quantized cyclotron orbit is outside the Mexican hat.
  (b) For certain $U$, the orbit goes inside the gap of the zero field spectrum.
  (c) At larger $U$, it moves underneath the Mexican hat where the direction of the group velocity is opposite to the momentum.
  (d) At very large $U$ (not presently accessible in experiments), where the BLG spectrum consists of two copies of monolayer spectra shifted by $\pm U$,
  the $n^{\rm th}$ electron Landau level of BLG approaches the $(n+1)^{\rm th}$ hole Landau level of the higher energy monolayer.
  \label{fig:BLG_LL_evolution}}
\end{figure}

The physical picture is as follows.
In a weak magnetic field, momentum ${\mib k}$ of a quasiparticle slowly rotates as a function of time $t$ according to the equation of motion
\begin{equation}
\dot{\Bk} = \omega_{\rm c} \, \zhat \times \Bk \quad,\quad
\omega_{\rm c} \equiv \dfrac{2 \pi }{ T} \sign(\vgr)\,,
\label{eqn:kdot}
\end{equation}
where $T = 2 \pi k_n \ell_{\!B}^2 / |\vgr(k_n)|$ is the cyclotron period.
(For simplicity, the valley and band labels are temporarily omitted.)
The rotation of ${\mib k}$ causes a slow evolution of the wavefunction $\ket{\alpha}$ in the pseudospin, i.e., sublattice space.
This causes the accumulation of the Berry phase~\cite{Berry1984qpi, Shapere1989gpp} 
\begin{equation}
\label{eqn:Berry_phase_def}
\PhB \equiv \sign(\vgr) \!\int_{0}^{T}\! dt \, \expect{\alpha}{i \Bnab_k}{\alpha} \cdot \dot{\Bk}\,.
\end{equation}
The quasiclassical quantization rule is~\cite{Chang1996bph}
\begin{equation} \label{eqn:scl_quan_cond}
\begin{split}
\sign(\vgr) \oint  d\pi_y \>\ell_{\!B}^2 \, \pi_x+ \PhB
&= \pi (k\ns_n \ellB)^{2}  + \PhB \\
&= (2 n + 1)\pi\,.
\end{split}
\end{equation}
This formula can be understood as a generalized Bohr-Sommerfeld rule: since $\ell_{\!B}^2\, \pi_x$ plays the role of ``momentum'' conjugate to the
``coordinate'' $\pi_y$, the top line represents the total phase shift acquired along the orbit, including the geometric phase.
Equation~\eqref{eqn:scl_quan_cond} establishes the  precise relation between the Onsager number $\delta$ and the Berry phase $\PhB$:
\begin{equation}
  \delta = \dfrac{1}{ 2} - \frac{\PhB}{ 2\pi}\,,
  \label{eqn:gamma}
\end{equation}
Thus, in monolayer graphene where $\PhB = \pi$, we get $\delta = 0$, which implies the existence of a level at zero energy.~\cite{Castroneto2009tep}

Comparing Eqs.~\eqref{eqn:D} and \eqref{eqn:Berry_phase_def} we see that for the isotropic spectrum, i.e., for $\gth = 0$, we have
\begin{equation}
\PhB =\dfrac{2 \pi k}{\vgr}\, D_\alpha \,.
\label{eqn:Phi_B_vs_D}
\end{equation}
Postponing the discussion of this equation for just a moment we note that for $\vgr \neq 0$, another version of the quantization rule can be established.~\cite{Gosselin2008sqo}
To this end one defines a modified orbit radius $\ktil_n$ such that  
\begin{equation}
\CE_{n,s\ns_1,s\ns_2} = E_{s\ns_1,s\ns_2, \ktil\ns_n}\,.
\label{eqn:E_quantized_II}
\end{equation}
To the leading order in $B$, the rule that determines $\ktil_n$ is similar to Eq.~\eqref{eqn:scl_quan_cond} except $\PhB$ is replaced by a different phase shift $\Phc\,$:
\begin{gather}
\pi \big(\ktil_n \ellB\big)^{2} = (2 n + 1)\,\pi - \Phc\,, \label{eqn:scl_quan_cond_II}\\
\Phc = \PhB + \dfrac{M B}{\hbar}\, T = \dfrac{\pi k}{\vgr} \big(F_\alpha + D_\alpha \big)\,.
\label{eqn:Phi_c}
\end{gather}
With further analysis it is possible to show that our $\Phc$ coincides with the ``semiclassical phase'' defined in Ref.~\onlinecite{Carmier2008bpi}.
Therefore, the difference between $\Phc$ and $\PhB$ noted in that paper is entirely due to the pseudo-Zeeman shift rather than a violation of adiabaticity.

Applying the above formulas to BLG, we obtain
\begin{equation}
\frac{\Phi_{\rm B}^{\pm}}{2\pi} = \mp \> \dfrac{U}{ 2 E_\alpha}
\left( 1 + s\ns_2 \dfrac{4 \ve_k^2 - \gamma_1^2}{\sqrt{\gamma_1^4 + 4 (\gamma_1^2 + 4 U^2)\, \ve_k^2}} \right).
\label{eqn:Berry_phase_final}
\end{equation}
At finite $U$ the Berry phase is a nonmonotonic function of momentum, which is addressed in more detail in Sec.~\ref{sec:AHE}.
Here we comment only on the simple case $U = 0$, where Eq.~\eqref{eqn:Berry_phase_final} gives $\Phi_{\rm B}^{\pm} = 0$ at all $k \neq 0$.
This seems to contradict to the assignment $\Phi_{\rm B}^{\pm} = \pm 2 \pi$ made in most of the previous work.~\cite{McCann2006lld, ZhangY2005eoo} In fact, there is no contradiction because the Berry phase is not unique: different choices for an overall phase of the wavefunction in Eq.~\eqref{eqn:Berry_phase_def} can shift $\Phi_{\rm B}^{\pm}$ by an arbitrary integer multiple of $2\pi$.
In the context of Landau quantization, such shifts can be compensated by relabeling the Landau index $n$, so that the physical quanitities --- the radii $k_n^{\pm}$ of the orbits and their energies --- remain the same. 

Combining Eqs.~\eqref{eqn:Phi_c} and \eqref{eqn:Berry_phase_final}, we obtain the analytic formula for the semiclassical phase:
\begin{equation}
\frac{\Phi_{\rm c}^{\pm}}{2\pi} = \mp \> s\ns_3\,
                 \dfrac{U E_\alpha}
                       {\sqrt{(\gamma_1^2 + 4U^2)E_\alpha^2 - \gamma_1^2\, U^2}} \,,
\label{eqn:sc_phase_final}
\end{equation}
This equation should be used away from momentum $\kstar$ where its denominator vanishes.
Finally, the quantization rule~\eqref{eqn:scl_quan_cond_II} becomes
\begin{equation}
  n + \half  = \bigg(\dfrac{\ve\ns_k}{\omo}\bigg)^{\!\!2} + \frac{\Phc}{ 2\pi}\,.
  \label{eqn:scl_quan_cond_III}
\end{equation}
In comparison, the precise relation between $n$ and $\CE_\alpha$ for the case $\gth = \gfo = \Dpr = 0$ reads
\begin{equation}
  n + \half = \dfrac{\CE_\alpha^2 + U^2 }{ \omega_0^2} -
  s\ns_3 \sqrt{ \dfrac{\sGamma^4(\CE\ns_\alpha)}{\omega_0^4} + \dfrac{2\,U\CE\ns_\alpha}{\omega_0^2} 
  + \dfrac{1}{ 4} \,}\,.
  \label{eqn:quan_cond}
\end{equation}
This result follows from Eq.~\eqref{purespec}; the composite label $\alpha$ denotes the set $\{n,s\ns_1,s\ns_2\}$.
The semiclassical Eq.~\eqref{eqn:scl_quan_cond_III} does agree with the exact Eq.~\eqref{eqn:quan_cond} to the leading order in $\omega_0^2$, i.e., ${\cal O}(1 / n)$ at large $n$.
Fortuitously, it is also valid for $n = -1$.
It predicts $\ktil_{1--}^+= 0$, which entails $\CE_{1--}^+= U$, in agreement with our earlier result.

The valley splitting of the Landau levels can be expressed as follows:
\begin{equation}
\CE_\alpha^+ - \CE_\alpha^- = -\dfrac{2 \hbar}{T}\, \Phi_{\rm c}^+
= s_1 s_2 \dfrac{2 \gamma_1^2 \beta U}{\sqrt{\gamma_1^4 + 4 (\gamma_1^2 + 4 U^2)\, \ve_k^2}}\,.
\label{eqn:valley-split}
\end{equation}
Here either $k_n^\pm$ or $\ktil_n^\pm$ can be used in place of $k$ because this formula is valid only to the leading order in $\beta$.
At low energies, it simplifies to
\begin{equation}
\CE_\alpha^+ - \CE_\alpha^- \simeq -2 \beta U\,, \quad n \gg 1\,,
\label{eqn:}
\end{equation}
in agreement with Eq.~\eqref{eqn:E_low_energy}.
We see that unlike the pseudo-Zeeman term, discussed in Sec.~\ref{sec:g_factor}, the net valley-splitting of the Landau levels has little energy or $n$ dependence.

It is now straightforward to apply the above quantization rules in order to understand qualitatively the evolution of some $n \gg 1$
Landau level as a function of $U$.
For the $K^+$ valley, illustrated in Fig.~\ref{fig:BLG_LL_evolution}, the situation is as follows.
As $U$ increases starting from zero, the radius of the orbit changes only slightly because ${\PhB} / {2\pi} \sim 1 \ll n$ for all $U$.
On the other hand, the Mexican hat expands in both height (energy) and width (momentum).
As a result, the quantized orbit slips from the exterior ($k > \kstar$) to the interior ($k < \kstar$) of the hat.
In the process, the orbit passes through a region where its energy is inside the gap of the $B = 0$ dispersion because of the negative pseudo-Zeeman term. [For the $K^+$ valley this occurs only if $U > 0$ but not if $U < 0$, see Eq.~\eqref{eqn:g_z_hat} and Sec.~\ref{sec:LL_spectrum} below.] Eventually, at very large $U$, the orbit approaches the $n + 1${st} hole Landau level of graphene monolayer, except it is shifted upward by $U$.

%%%%%%%%%%%%%%%%%%%%%%%%%%%%%%%%%%%%%%%%%%%%%%%%%%%%%%%%%%%%%%%%%%%%%%%%%%%%%%%
\section{Landau level spectrum}
\label{sec:LL_spectrum}

%%%%%%%%%%%%%%%%%%%%%%%%%%%%%%%%%%%%%%%%%%%%%%%%%%%%%%%%%%%%%%%%%%%%%%%%%%%%%%%
\subsection{Level crossings}
\label{Sec:LL_crossing}
%%%%%%%%%%%%%%%%%%%%%%%%%%%%%%%%%%%%%%%%%%%%%%%%%%%%%%%%%%%%%%%%%%%%%%%%%%%%%%%

In this section we explain the physical origin of a nonmonotonic $U$-dependence of Landau level energies, which gives rise to a complicated net-like pattern with numerous crossings, see Figs.~\ref{fig:vspecc} and \ref{fig:vspeca}.
It should be clarified that electron interactions, which are ignored in our calculations, can produce significant corrections to the Landau level spectrum.
However, we expect that topological properties of the level diagram would not change much.

Figure~\ref{fig:vspecc} shows the first several Landau levels, which we calculated numerically as a function of $U$ at a representative magnetic field of $B = 5\,\text{T}$. Only $U > 0$ are shown because the energies at negative $U$ can be obtained from the symmetry relation $\CE^+_\alpha(U) = \CE^-_\alpha(-U)$. Let us focus on the $s\ns_1 = +1$ levels and consider the limits of small and large $U$ (a similar argument can be applied to the $s\ns_1 = -1$ levels with appropriate sign changes).

For small $U$, the Landau levels are roughly equidistant and those with higher index $n$ have higher energies
(in agreement with Eq.~\eqref{purespec} for $U = 0$).
In the opposite limit of $U \gg \gon$, from Eq.~\eqref{HpB} it is easy to see that the BLG spectrum consists of two copies of the monolayer spectrum shifted by $\pm U$.
Accordingly, the set $\big\{\CE^\pm_{n+-}\big\}$ approaches the Landau level energies of the holes in the monolayer,~\cite{Castroneto2009tep} but shifted by $U > 0$:
\begin{equation}
\CE^+_{n+-}\simeq U - \sqrt{n + 1}\> \omega_0 \quad,\quad
\CE^-_{n+-} \simeq U - \sqrt{n}\>\omega_0 \,.
\label{eqn:E+_large_U-}
\end{equation}
In this limit states of higher index have lower energies.
Therefore, any two levels of the $s\ns_2 = -1$ band cross at some value of $U$.
This occurs when the corresponding quantized orbits are located at the same energy but on the opposite sides of the Mexican hat
(see Fig.~\ref{fig:BLG_LL_evolution}). 

In addition, it is possible to have crossings of orbits on the same side of the Mexican hat if they belong to opposite valleys.
In the semiclassical approximation, this occurs whenever $\Phi_{\rm c}^{\pm} / \pi$ is an integer.
In this case the difference in $n$ is compensated by the difference in the semiclassical phase, yielding the same momentum $\ktil_n$ and energy $E_{s\ns_1,s\ns_2}(\ktil_n)$ (see Eq.~\eqref{eqn:scl_quan_cond_II}).
For example, at $U = 0$ we have $\Phi_{\rm c}^{\pm} = 0$, so that all
Landau levels should be (and are) valley-degenerate.
Next, $|\Phi_{\rm c}^{\pm}| \to \pi$ as $U \to \infty$, so in this limit the \emph{adjacent} Landau levels coincide, in agreement with Eq.~\eqref{eqn:E+_large_U-}.
Using Eqs.~\eqref{eqn:E_*}, \eqref{eqn:Gamma}, and \eqref{eqn:sc_phase_final}, one can show that the condition $|\Phi_{\rm c}^{\pm}| = N \pi$ is met at
\begin{equation}
 {\cal E}_\alpha^2 = \dfrac{E_\star^2 }{ 1 - (2 \Estar / N \gon)^2} \qquad
 (\gth = \gfo = \Dpr = 0)\,.
\label{eqn:integer_Phi_c}
\end{equation}
This implies that the level crossings are confined to the range of energies $\Estar \leq |\CE_\alpha| \leq |U|$, which is precisely the range between the top and the bottom of the Mexican hat.
The crossings at the top of the hat are between the adjacent Landau levels ($N = 1$).
Since the special level $\CE^\pm_{-1} = \pm U$ also happens to be at the same energy, these are actually {\it triple} crossings.
In the $s\ns_1 = \pm 1$ band, they involve $n${th} level of $ K^\pm$, the $n - 1${st} level of $K^\mp$, and the $-1$ level of $K^\pm$ (assuming $U > 0$).
When $\gth=\gfo=\Dpr=0$ these unusual triple crossings appear when $U=U_n \equiv \half\sqrt{n} \> \omo$.
We can show by algebraic means that finite $\gfo$ and $\Dpr$ give corrections to $U_n$ but do not lift the triple degeneracies.
We suspect that this property stems from some hidden symmetry of the Hamiltonians $\CH_n^\pm$.

%%%%%%%%%%%%%%%%%%%%%%%%%%%%%%%%%%%%%%%%%%%%%%%%%%%%%%%%%%%%%%%%%%%%%%%%%%%%%%%
\subsection{Trigonal warping}
\label{sec:trigonal}
%%%%%%%%%%%%%%%%%%%%%%%%%%%%%%%%%%%%%%%%%%%%%%%%%%%%%%%%%%%%%%%%%%%%%%%%%%%%%%%

The parameter $\gth$ has a number of interesting effects on both the zero-field and Landau level spectra.
It mixes Landau levels of the same valley with indices $n$ different by an integer multiple of three, see Appendix~\ref{sec:BLG_H_n}.
This turns crossings between such levels into avoided crossings.
Strictly speaking, we can no longer label Landau levels by $\{n,s\ns_1,s\ns_2\}$.
Nevertheless, the effect of $\gamma\ns_3$ is small at low $U$, so that with proper care it is possible to track the levels through the avoided crossings and still retain our labeling scheme.
The calculation of the
Landau level spectra with $\gth \neq 0$ is handled numerically.
To account for the level mixing at high $U$ we had to diagonalize matrices of size $4J$ with large enough $J$ ($J \approx100$) to ensure numerical accuracy, see Appendix~\ref{sec:BLG_H_n}.
One effect of $\gth$ is to lift the triple degeneracy of the adjacent Landau levels by moving the crossing point energy away from the top of the
Mexican hat, as expected.

\begin{figure}[t]
  \begin{center}
    \includegraphics[width=3.2in]{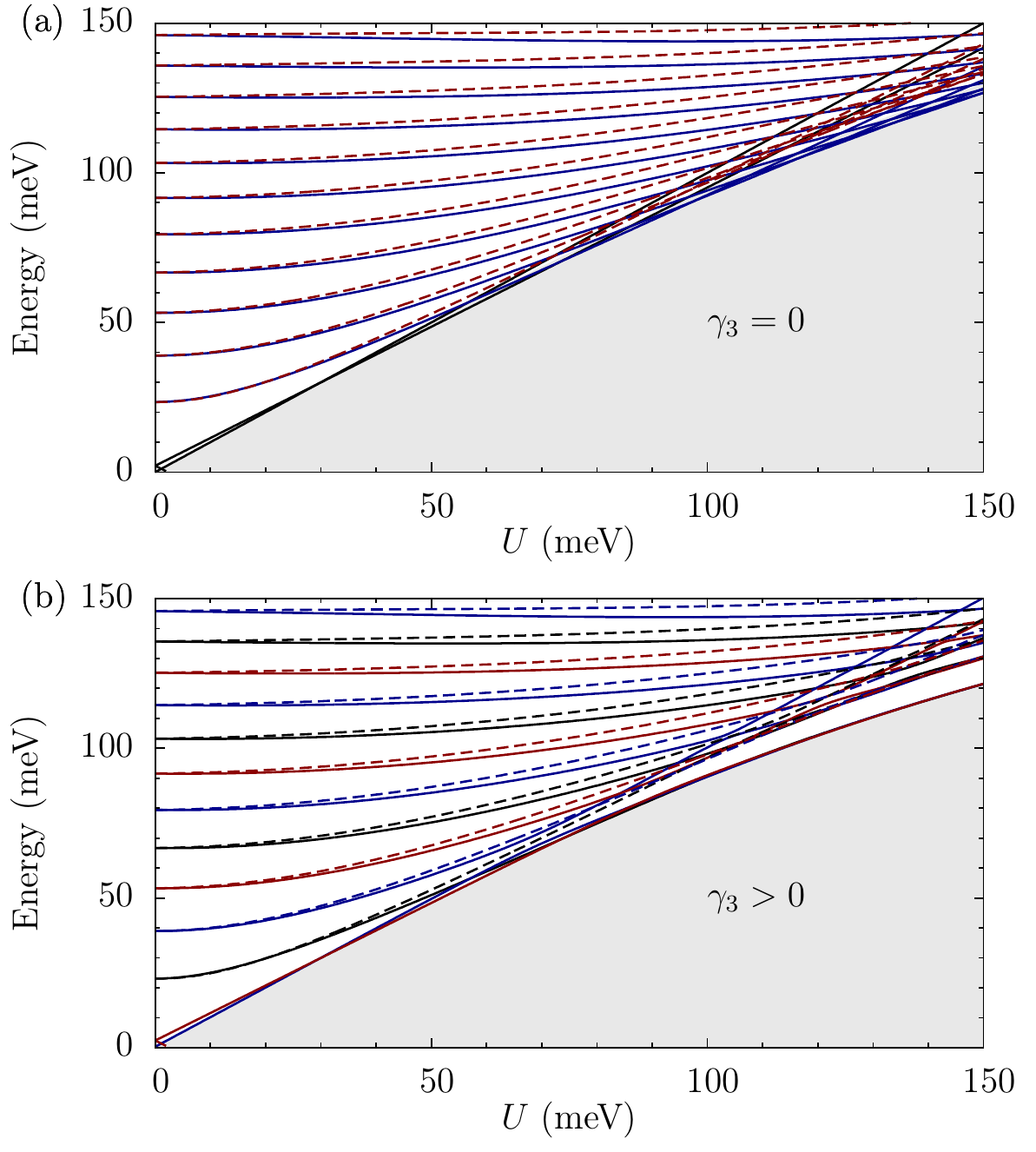}
  \end{center}
  \caption{(Color online) Landau level energies {\it vs.} interlayer bias $U$ for a field value $B=5\,$T.
  (a) Top panel: $\gamma_3=0$; (b) bottom panel: $\gamma_3=0.3\,$eV.
  The color and line type are as in Fig.~\ref{fig:vspecb}.
  Note the bunching of levels at the edges of the central band gap when $\gamma_3\ne 0$: the two levels just below the gap for $U \gtwid 100\,$meV are both very nearly threefold degenerate.}
  \label{fig:vspecc}
\end{figure}

A more interesting effect is the shift of the $B = 0$ band edges, which are the boundaries of the central band gap in Fig.~\ref{fig:vspecb}.
This can be understood as follows.
The hopping $\gth$ induces a trigonal warping of the zero-field bands, as described by
Eq.~\eqref{eqn:b_square}.
Accordingly, the low-energy region of the conduction band develops three kidney-shaped pockets along $k = \kstar$ circle centered, in $K^+$ valley, at $\vphi =\frac{1}{3}\pi$, $\pi$, and $\frac{5}{3}\pi$ angular positions.~\footnote{This agrees with Fig.~6 of
Ref.~\onlinecite{Nilsson2008epo} where $\gamma_3  > 0$ is also assumed.
In contrast, in Ref.~\onlinecite{McCann2006lld} where the sign of $\gamma_3$ is opposite to ours, values of $\vphi$ are shifted by $\pi$.}
To the leading order in $\gth$, their energy is lowered below $\Estar$ by
\begin{equation}
\delta E \simeq \dfrac{\sqrt{8}\,\gth}{\gze}\, \dfrac{U^2}{\gon} \qquad
\left(\frac{\gth}{\gze} \ll \frac{U}{\gon} \ll 1\right)\,,  \label{eqn:dE3_small_B}
\end{equation}
which follows from Eqs.~\eqref{eqn:b_square}.
Accordingly, the band edge of the conduction/valence band at $B = 0$ shifts by $\mp\delta E$.
For example, at $U = 0.15\,\text{eV}$ we obtain $\delta E \approx 8\,\text{meV}$.
This is in a good agreement the numerical results shown in
Figs.~\ref{fig:vspecb} and \ref{fig:vspecc}.

%%%%%%%%%%%%%%%%%%%%%%%%%%%%%%%%%%%%%%%%%%%%%%%%%%%%%%%%%
%%%%%%%%%%%%%%%%%%%%%%%%%%%%%%%%%%%%%%%%%%%%%%%%%%%%%%%%%

\begin{figure}[b]
\centering
\includegraphics[width=3.0in]{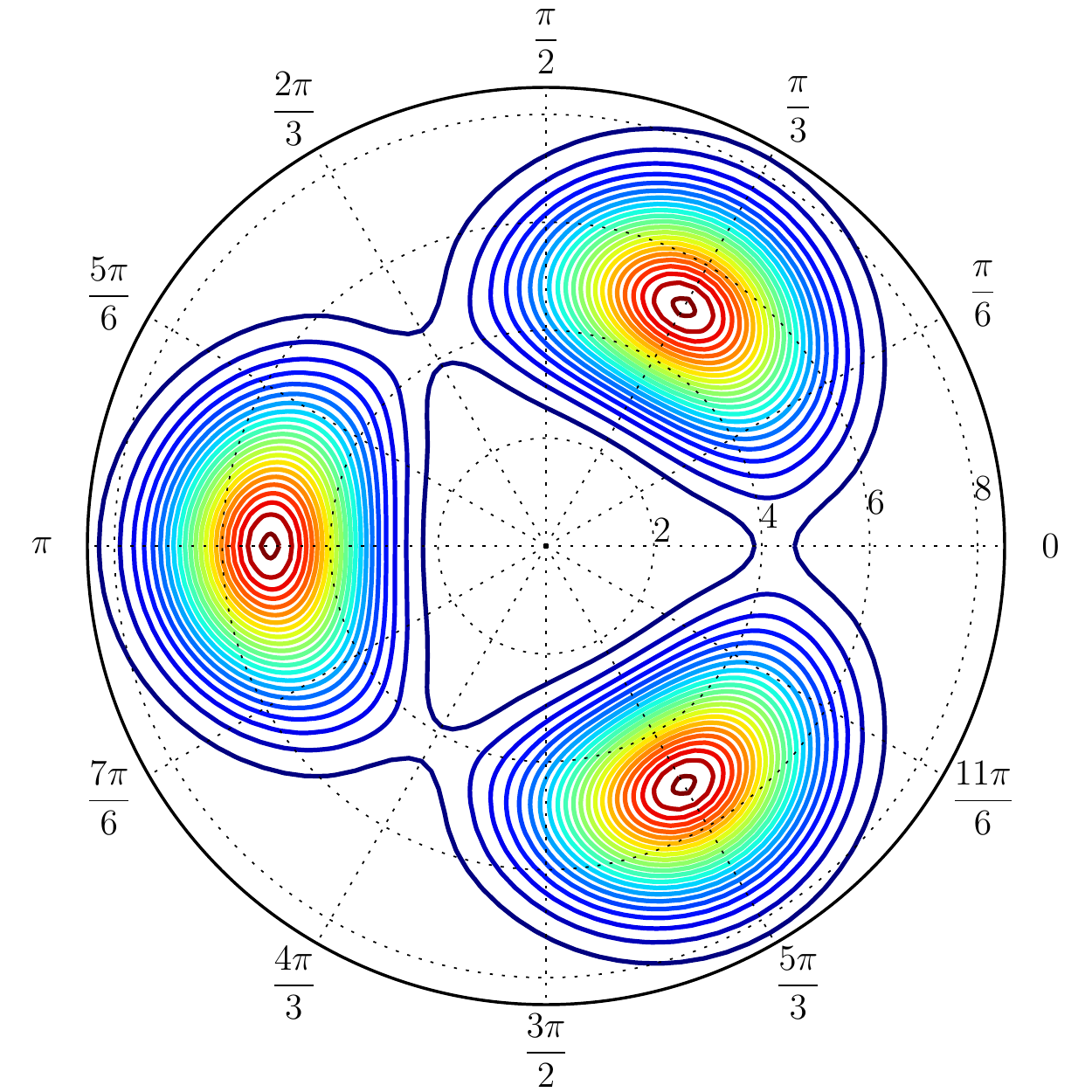}
\caption{(Color online) Absolute value of the Bloch function for the lowest-energy Landau level of the conduction band.
The origin is at the $K^+$ point, the radial coordinate is $k {\ell_B}$, and $U = 0.15\,\text{eV}$.}
\label{fig:pockets}
\end{figure}

The effect of $\gamma\ns_3$ on Landau levels is even more striking.
As one can see from Fig.~\ref{fig:vspecc}, it leads to a bunching of Landau levels near the conduction (and valence) band edges as $U$ increases above $0.1\,\text{eV}$.
Apparently, these Landau levels, which can be labeled $n = n\ns_\star - 1$, $n\ns_\star$, and $n\ns_\star + 1$, become nearly degenerate.
Within a simple quasiclassical picture, the explanation is straightforward: this trio of levels correspond to three orbits, which are identical in shape and energy but are separately confined inside the three equivalent pockets.~\cite{McCann2006lld} In a more refined description, such orbits are hybridized by a weak quantum tunneling, so that the Bloch functions have equal amplitude in each pocket but different phases.
To verify this picture, we chose a set of $U$ in the range between $0$ and $0.15\,\text{eV}$ and for each of them computed the Bloch function of the lowest-energy state numerically.
We took $\gth=0.15$, for which there is only a single threefold degenerate level lying just within the central gap.
At all $U$, these functions exhibit maxima centered at $\varphi =\frac{1}{3}\pi$, $\pi$, and $\frac{5}{3}\pi$, as expected (see Fig.~\ref{fig:pockets}).
However, for $U \gtwid 0.1\,\text{eV}$ such maxima become very sharp, consistent with the picture of confinement and in concert with the coalescence of the energy levels into a single narrow bunch, as in Fig.~\ref{fig:vspecc}.

%%%%%%%%%%%%%%%%%%%%%%%%%%%%%%%%%%%%%%%%%%%%%%%%%%%%%%%%%
%%%%%%%%%%%%%%%%%%%%%%%%%%%%%%%%%%%%%%%%%%%%%%%%%%%%%%%%%

%n Fig.~\ref{fig:BLG_LL_Vb}(b) the energies of this bunch in the conduction and its counterpart in the valence band seem to be lined up with
%the respective edges of $B = 0$ spectrum, as though the pseudo-Zeeman effect is canceled.
%This cancellation is fortuitous.
%We attribute it to the zero point motion of the orbits confined inside the pockets.
%Clearly, the Bloch functions (Fig.~\ref{fig:pockets}) have some finite spread around the centers of the pockets.
%Thus, in the conduction band such orbits are raised in energy above the actual minima of the band, which counteracts the effect of the
%pseudo-Zeeman shift.
%Indeed, a better measure of the pseudo-Zeeman effect is the valley splitting, which is nearly the same in
%Fig.~\ref{fig:BLG_LL_Vb}(a) and \ref{fig:BLG_LL_Vb}(b).
%The magnitude of the zero-point energy shift depends on $U$ and $B$ and
%just happens to be numerically close to the pseudo-Zeeman shift in a range of $U$ shown in Fig.~\ref{fig:BLG_LL_Vb}(b).

In general, the influence of $\gth$ on the spectrum gets stronger as $B$ decreases or $U$ increases.
This is because the depth $\delta E$ of the pockets and their width increases with $U$ while the area in momentum space per orbit is equal to $2\pi / \ell_{\!B}^2 \propto B$, as discussed in Sec.~\ref{sec:quasiclassical}.
Hence, at large $U$ and/or small $B$, each pocket may host several orbits, so that higher-energy Landau levels can also form bunches of three, as is apparent in Fig.~\ref{fig:vspecc}, where there are two nearly three-fold degenerate sets of Landau levels separated by 10-20 meV from a tangle of higher energy states.
The first bunch emerges at $U\approx 80\,$meV and the second at $U\approx 120\,$meV.
Conversely, as $B$ increases at fixed $U$, separate orbits no longer fit into the pockets and they unite into a single contiguous loop.
At this point, the effect of $\gth$ can safely be neglected.

% The change to the gap energy introduced by $\gamma_3$ is of order $B^{-2}$:
% %%
% %%
% \begin{equation}
%   \delta E_{\gamma_{3}} \simeq
%   - v\ns_3^2 \frac{U \gamma_1^4 \hbar^2 c^2}{3 e^2 B^2}
%   + O(B^{-3}) \,
%   \label{eqn:dE3_large_B}
% \end{equation}
% %%

\subsection{Energy gap}
%%%%%%%%%%%%%%%%%%%%%%%%%%%%%%%%%%%%%%%%%%%%%%%%%%%%%%%%%%%%%%%%%%%%%%%%%%%%%%%

The above discussion indicates that the energy gap of BLG can be controlled not only by $U$ but also by $B$ while keeping $U$ fixed.
Since this gap can strongly affect the low-temperature transport, it may be of interest in applications, and so it deserves some discussion.
The magnetic field tends to reduce the gap relative to the zero-field case, as one can see in Figs.~\ref{fig:vspecb} and \ref{fig:vspecc}, where the gray area indicates the zero field gap.
In other words, some Landau levels can reside \emph{inside} the bandgap $|E| < \Estar$ of the $B = 0$ spectrum.
This phenomenon is a direct manifestation of the pseudo-Zeeman shift.
It is seen more clearly in Fig.~\ref{fig:BLG_LL_evolution}(b), where only one
Landau level (from the $K^+$ valley) is shown.
For a certain $U > 0$ this level drops below the zero-field minimum $\Estar$ of the conduction band.
Similarly, there is another Landau level from $K^+$ valley, not shown in the Fig.~\ref{fig:BLG_LL_evolution}(b), which rises above the maximum $-\Estar$ of the valence band.
This is because the pseudo-Zeeman effect has opposite signs in the two valleys.
Based on this argument, we can use Eq.~\eqref{eqn:g_z_hat} to show that, e.g., the bottom of the conduction band shifts to
\begin{equation}
\label{eqn:Eg_in_B}
\Etil\ns_\star \simeq  \Estar - \dfrac{\beta\,|U|}{1 + (2 U / \gon)^2} \,,
\end{equation}
where $\beta$ is defined in Eq.~\eqref{eqn:beta}.
In principle, this approximate formula can be refined by semiclassical quantization.
The true band edge is determined by the lowest-energy Landau level of the conduction band.
Its index $n\ns_\star$, which depends on $U$ and $B$, can be found by setting $\ktil = \kstar$ and dropping the second term on the right-hand side of
Eq.~\eqref{eqn:scl_quan_cond_III}: $n\ns_\star + \half \simeq {\ve_\star^2} / {\omega_0^2} = (1 / \beta) (\vestar / \gamma\ns_1)^2$.
A similar result can be obtained from the low-energy effective theory, by minimizing the energy in Eq.~\eqref{eqn:E_low_energy} with respect to the Landau index $n$. With $\Dpr = 0$ we obtain
\begin{equation}
  n\ns_\star=\dfrac{1}{ \beta}\,\dfrac{2U^2}{ \gamma_1^2+4U^2}\,.
  \label{eqn:n_*}
\end{equation}
%%

%%%%%%%%%%%%%%%%%%%%%%%%%%%%%%%%%%%%%%%%%%%%%%%%%%%%%%%%%%%%%%%%%%%%%%%%%%%%%%%
\begin{figure}[t]
  \centering
  \includegraphics[width=3.3in]{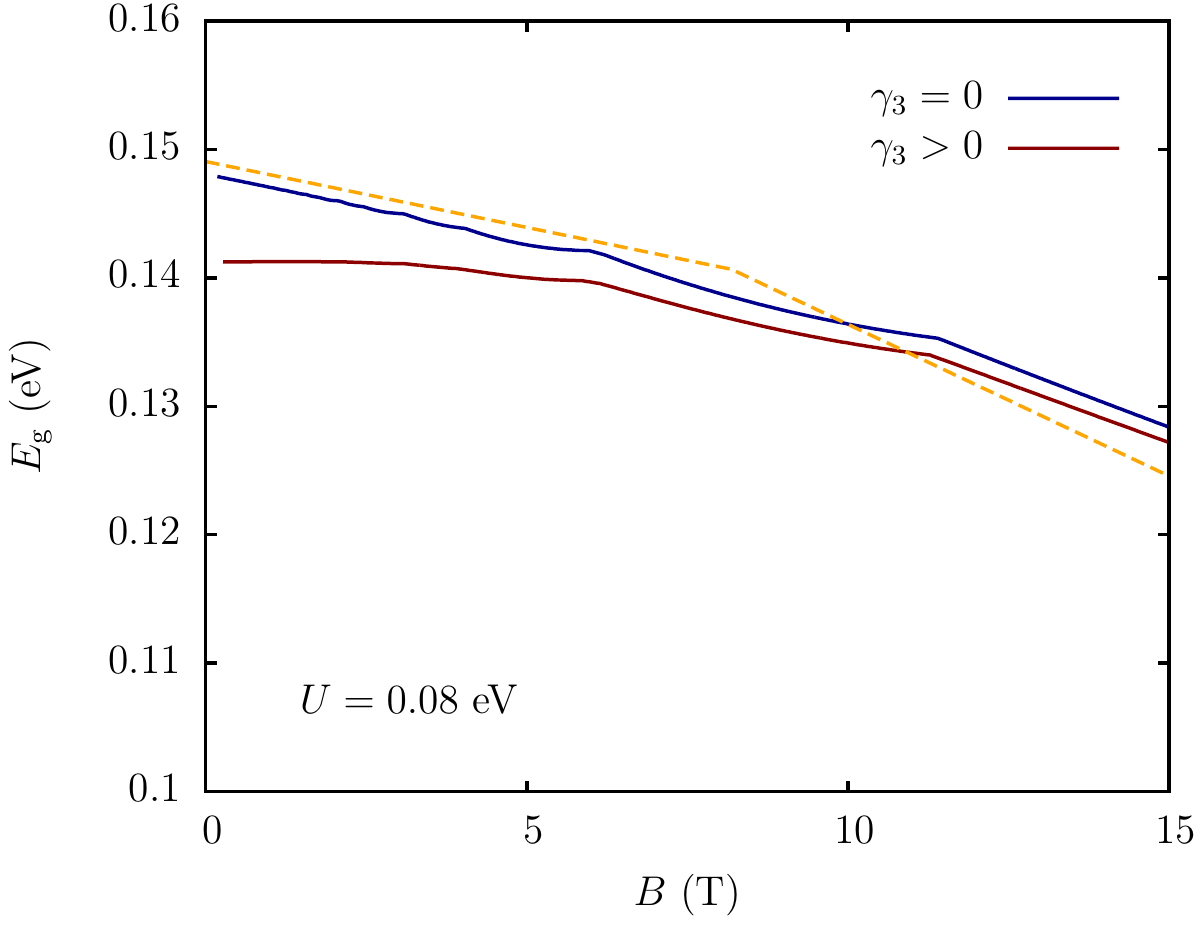}
  \caption{(Color online) Energy gap separating Landau levels of the valence bands from those of the conduction band as a function of the magnetic field $B$.
  The cusps on the curves are due to discrete changes in Landau level index $n_\star$ (see main text).
  The upper solid curve is for
  $\gamma_3 = 0$ and  the lower one for $\gamma_3 = 0.30\,\text{eV}$.
  The analytic estimate per Eqs.~\eqref{eqn:Eg_in_B} and
  \eqref{eqn:Eg_in_large_B} is shown by the dashed line.}
  \label{fig:BLG_gap}
\end{figure}
%%%%%%%%%%%%%%%%%%%%%%%%%%%%%%%%%%%%%%%%%%%%%%%%%%%%%%%%%%%%%%%%%%%%%%%%%%%%%%%

Since $\beta \propto B$, our approximate formula $2 \Etil_\star^+$ for the gap predicts a linear gap narrowing as $B$ increases at $U = \text{const}$.
Figure~\ref{fig:BLG_gap} demonstrates that it is quite accurate up to the point where $n_{\dagger}$ drops to zero, i.e., up to the field where $\beta \approx 2 U^2 / \gamma_1^2$.
Of course, this approximation misses the small cusps produced by the discrete changes in $n\ns_\star$.

At larger $B$, the gap is determined by the energy of the special $n = 0$ Landau level for which Eq.~\eqref{eqn:Eg_in_B} is not valid.
Instead, we can use Eq.~\eqref{eqn:E_0} to get
\begin{equation}
\Etil\ns_\star \simeq E\ns_\star - \big(2 |U| - \Dtil\big)\,\beta\,,
\qquad \beta < 2U^2/\gamma_1^2\,.
\label{eqn:Eg_in_large_B}
\end{equation}
We see that the $B$-dependence remains linear but the slope becomes larger by a factor of two or so.
This prediction is in a reasonable agreement with numerical calculations (Fig.~\ref{fig:BLG_gap}).
The deviations seen at $B \gtrsim 10\,\text{T}$ are due to insufficient accuracy of the low-energy theory at such fields.
The total reduction of the gap as the field changes from $B = 0\,\text{T}$ to $15\,$T is about $15\,\text{meV}$ or $10\%$.

At even larger $B$, level $n = 0$ on the $s\ns_1 = \sign(U)$ side would cross with level $n = -1$, so that the slope of the linear dependence would change again.
That $n = -1$ level would eventually intersect with the other $n = 0$ level if $B$ keeps increasing, at which point the gap would momentarily vanish.
An example of such an intersection is shown in Fig.~\ref{fig:BLG_LL_Vb_zoom} (although the energies are plotted as a function of $U$).

%For completeness, we mention that the gap would vanish at the field
%$B_{{{\dagger}}} = ({\hbar c} / {2e}) (\gamma_1^2 + U^2)$, which is around
%$130\,\text{T}$.
%Needless to say this $B$ is difficult to reach in practice.
%With further increase of $B$, the gap would reappear and in the limit $B \to \infty$, the energy of $n = 0$ level would approach $-U$.

Let us now discuss the effect of $\gth$.
In Fig.~\ref{fig:vspecb}, the energies of the lowest-energy levels of the conduction and its counterpart in the valence band seem to be lined up with the respective edges of the $B = 0$ spectrum, as though the pseudo-Zeeman effect is canceled.
This cancellation is fortuitous.
We attribute it to the zero point motion of the orbits confined inside the pockets.
Clearly, the Bloch functions (Fig.~\ref{fig:pockets}) have some finite spread around the centers of the pockets.
Thus, in the conduction band such orbits are raised in energy above the actual minima of the band, which counteracts the effect of the pseudo-Zeeman shift.
Indeed, a better measure of the pseudo-Zeeman effect is the valley splitting, which is nearly the same in Figs.~\ref{fig:vspecc}(a) and (b). 
(The latter is essentially the upper half of Fig.~\ref{fig:vspecb}).
The magnitude of the zero-point energy shift depends on $U$ and $B$ and just happens to be numerically close to the pseudo-Zeeman shift in a range of parameters used in Fig.~\ref{fig:vspecb}.

The gap narrowing becomes more apparent at larger magnetic fields, see Fig. 12.
The upper and the lower solid curves represent the energy gap without and with $\gamma\ns_3$, respectively.
At $B = 0$, the distance between the two curves is about $8\,\text{meV}$, which is close to $2\, \delta E \approx 9\,\text{meV}$ per Eq.~\eqref{eqn:dE3_small_B}.
As $B$ increases, this distance quickly diminishes, and the gap concomitantly narrows.

%%%%%%%%%%%%%%%%%%%%%%%%%%%%%%%%%%%%%%%%%%%%%%%%%%%%%%%%%%%%%%%%%%%%%%%%%%%%%%%
\section{Anomalous Hall conductivity and valley magnetization}
\label{sec:AHE}
%%%%%%%%%%%%%%%%%%%%%%%%%%%%%%%%%%%%%%%%%%%%%%%%%%%%%%%%%%%%%%%%%%%%%%%%%%%%%%%

Systems that carry a finite Berry phase can exhibit a nonzero Hall conductivity $\sigH$ even at $B = 0$.
This is known as the anomalous Hall effect (AHE).
The AHE and other manifestations of the Berry phase in electronic properties have been recently reviewed in Ref.~\onlinecite{Xiao2010bpe}.
It has been shown that for a partially filled band, $\sigH$ in units of $e^2 / h$ is equal to the Berry curvature
\begin{equation}
\Omega \equiv [\Bnab\ns_k \times \expect{\alpha}{i \Bnab\ns_k}{\alpha}] \cdot \zhat\,,
\label{eqn:Berry_curvature}
\end{equation}
integrated over all occupied states.
By the Stokes' theorem, in the two-dimensional case the result is determined solely by the Berry phase at the Fermi level.
Therefore, we can readily compute the anomalous contribution to $\sigH$ from our Eq.~\eqref{eqn:Berry_phase_final}.
To do so, we need the Berry phase as a function of energy.
Substituting Eq.~\eqref{eqn:GamLam} into Eq.~\eqref{eqn:Berry_phase_final}, we obtain
\begin{equation}
\dfrac{\Phi_{\rm B}^\pm}{2\pi} = \pm\dfrac{U}{ E} \,
\dfrac{2E^2 - \gamma_1^2 - 2s_3 \sGamma^2(E)}
      {4U^2 + \gamma_1^2 - 2s_3 \sGamma^2(E)} \,,
\label{eqn:sig_H_aux}
\end{equation}
where $\sGamma(E)$ is defined in Eq.~\eqref{eqn:Gamma}.
The opposite signs in this formula indicate that the two valleys give opposite contributions to the AHE.
Therefore, $\sigH$ is nonzero only if unequal population of the valleys is created.
While this occurs naturally for $B\ne 0$, we desire, in the context of the AHE, that it should also occur in the absence of an external magnetic field.
Theoretical proposals for achieving that have been advanced in Refs.~\onlinecite{Rycerz07vvg, Xiao2007vpg, Yao2008prb, Moskalenko2009liv}.
Here we do not address any mechanisms of valley polarization but simply compute all the quantities for $K^+$ valley only.
Comparison with previous work will be given at the end of this section.

\begin{figure}[b]
  \begin{center}
  \vbox{
    \includegraphics[width=3.0in]{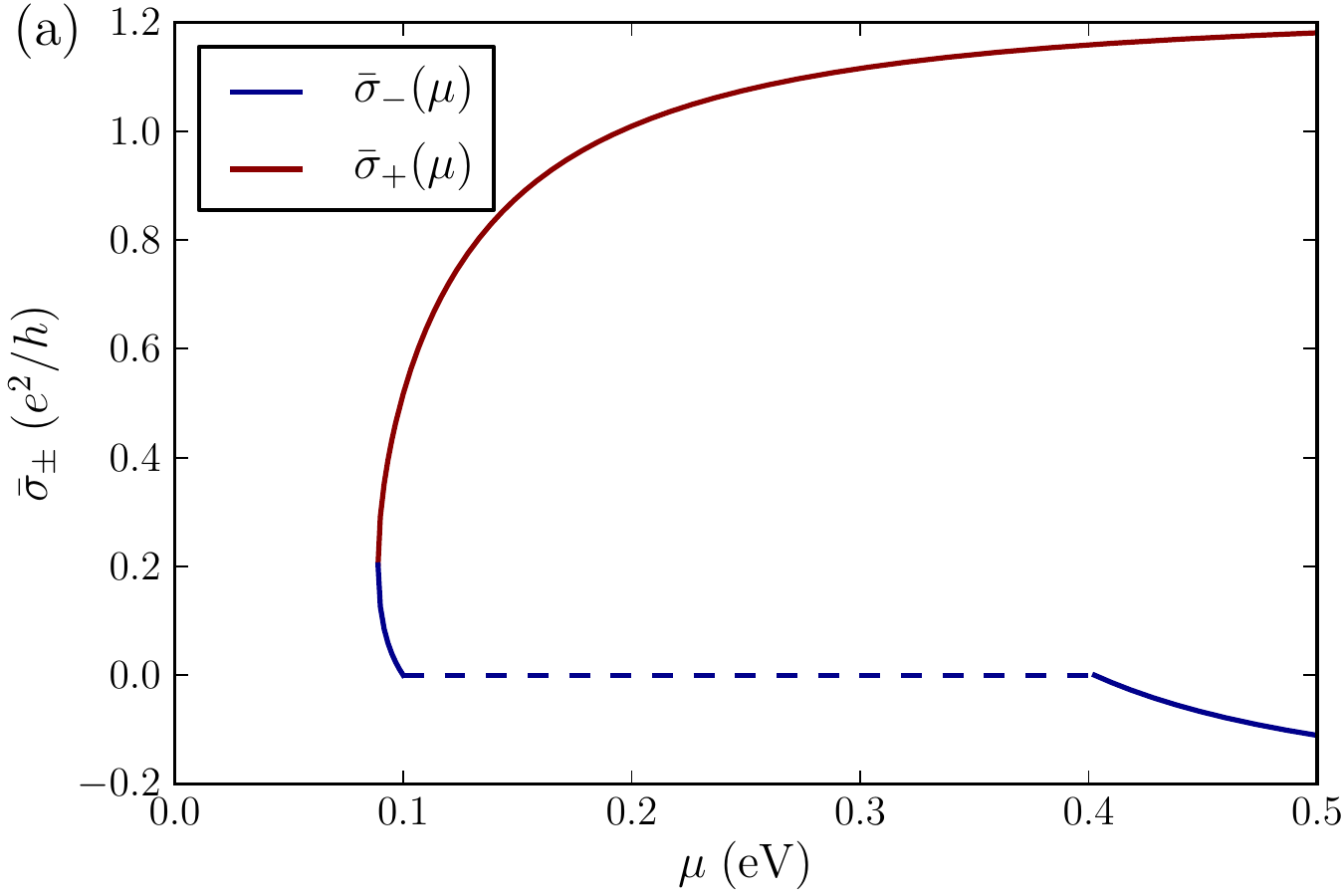}\\
    \includegraphics[width=3.0in]{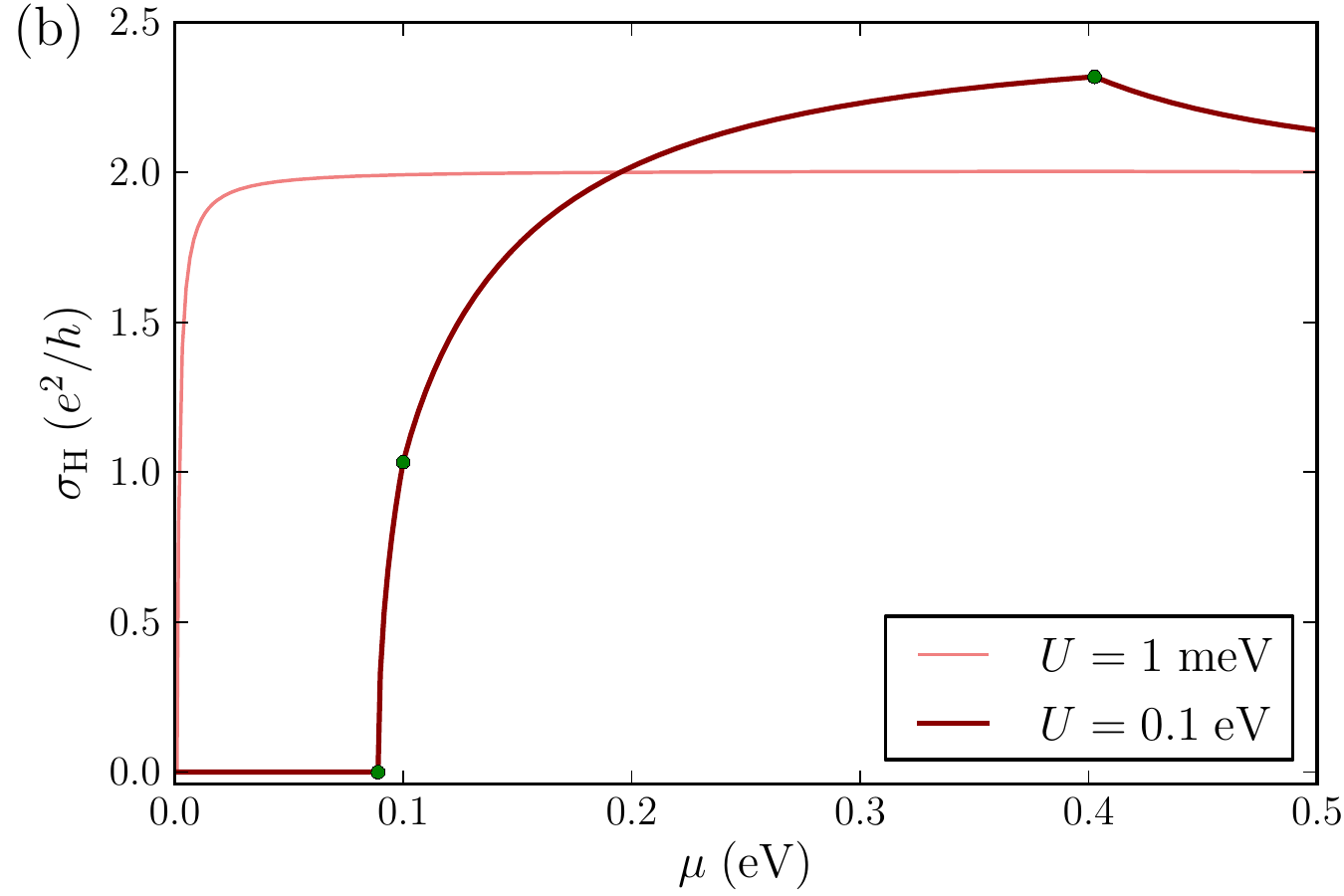}\\
    \includegraphics[width=3.0in]{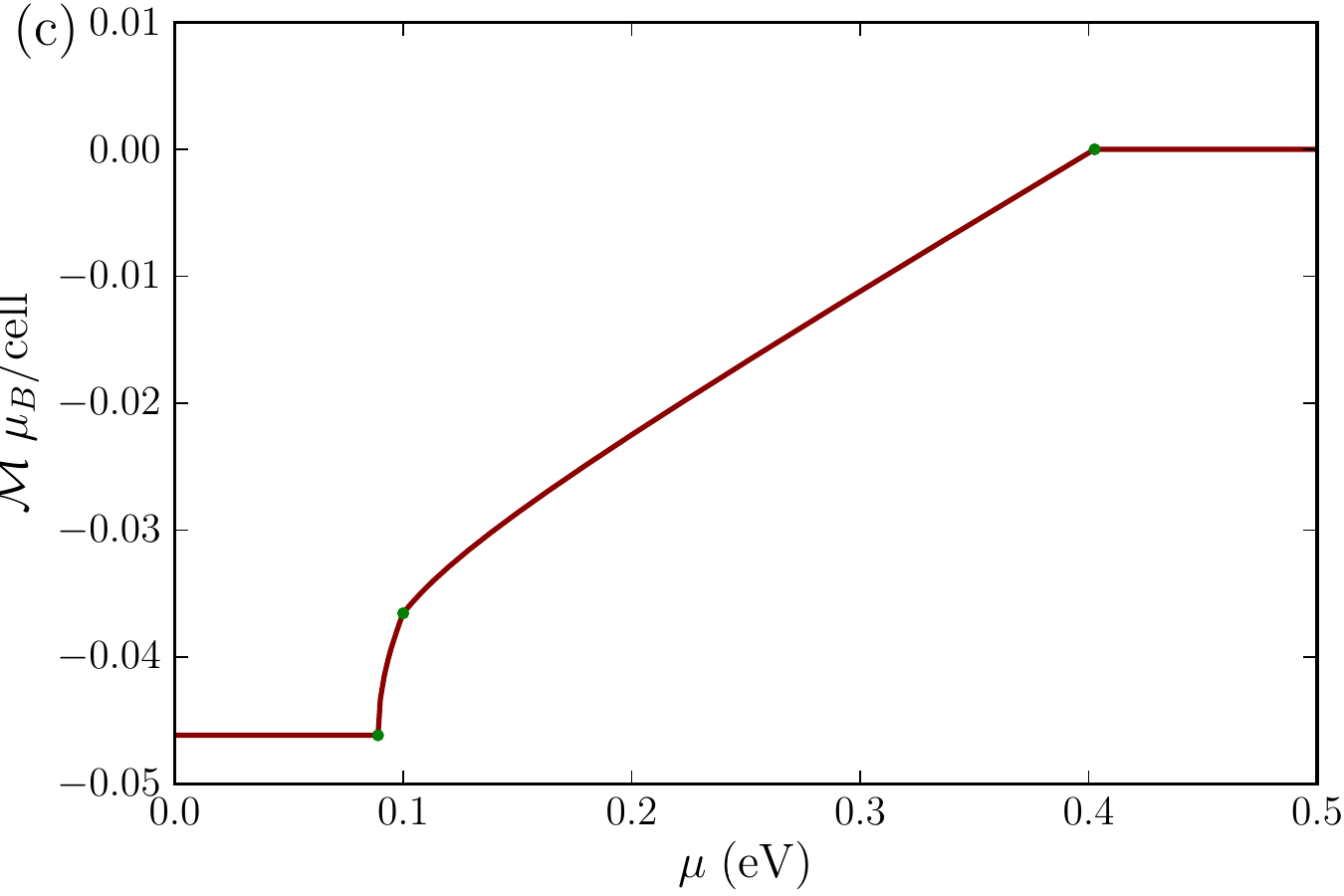}
  }
  \end{center}
  \caption{(Color online) At energy $|\mu| > E_{{{\dagger}}}$, there can be one or two branches of $k(\mu)$, each with its own
  (a) contribution to the anomalous Hall conductivity.
  As a result the (b) total anomalous Hall conductivity has cusps where Fermi surface's topology changes.
  The same also occurs for the (c) Magnetization density ${\cal M}(\mu)$ as function of chemical potential $\mu$.
  The solid curves in each are for $U \approx 0.1 \textrm{eV}$.  $\sigma^{\ }_{\rm H}$ for $U \approx 1$ meV is also shown in (b) as
  dashed trace that it approaches the universal value of 2.
  The negative value for ${\cal M}(0)$ is due to contribution from the two filled
  valence bands.
  When contribution from all four bands are included, ${\cal M}$ eventually adds to zero.} 
  \label{fig:M_mu}
\end{figure}

For brevity, we limit the consideration to the case where the Fermi level $\mu$ resides in the conduction bands ($s_1 = +$), i.e., $\mu > 0$.
Using Eq.~\eqref{eqn:Berry_phase_final}, we obtain:
\begin{equation}
  \sigH = \begin{cases} 
    0 & \mu < \Estar \,,\\
    \bar{\sigma}_-(\mu) - \bar{\sigma}_+(\mu) & \Estar \le \mu < U \,,\\
    \bar{\sigma}_-(\mu) - \bar{\sigma}_+(U) & U \le \mu < E\ns_\diamond \,,\\
    \bar{\sigma}_-(\mu) - \bar{\sigma}_+(U) + \bar{\sigma}_{+}(\mu) & E\ns_\diamond \le \mu \,,
  \end{cases}\label{eqn:sig_H}
\end{equation}
where
\begin{equation}
\bar{\sigma}\ns_{s\ns_3}(E)\equiv \frac{g\ns_{\rm s} e^2}{h}\,\Phi_{\rm B}^+ (s\ns_3)\,,
\end{equation}
$g_{\rm s} = 2$ is the spin degeneracy, and
\begin{equation}
E\ns_\diamond \equiv E\ns_{++}(k=0) = \sqrt{\gamma_1^2 + U^2} \,.
\label{eqn:E_ss}
\end{equation}
Hence, $\Estar$, $U$, and $E\ns_\diamond$ are the energies of the Mexican hat bottom, Mexican hat top, and the upper conduction band bottom, respectively.
At these energies the topology of the Fermi surface changes: from two concentric circles to one and back to two  (we ignore $\gth$).
Accordingly, $\sigH(\mu, U)$ is nonanalytic at such energies: it has discontinuous derivative (cusps), which are marked by the dots in
Fig.~\ref{fig:M_mu}(a).
Note that in the limit of small $U$, $\sigH$ approaches the universal value of $2e^2/ h$.
This property is related to the ``double step'' of the usual Hall conductivity at zero density (electroneutrality), which is a hallmark of the quantum Hall effect in a symmetric graphene bilayer.~\cite{Castroneto2009tep}

%%
%%%%%%%%%%%%%%%%%%%%%%%%%%%%%%%%%%%%%%%%%%%%%%%%%%%%%%%%%%%%%%%%%%
%%%%%%%%%%%%%%%%%%%%%%%%%%%%%%%%%%%%%%%%%%%%%%%%%%%%%%%%%%%%%%%%%%
%% 

Another quantity we can easily compute is the total magnetization ${\cal M}$ of the $K^+$ valley.
Recall that at finite $U$ each state $\{s\ns_1,s\ns_2,\Bk\}$ carries the orbital magnetic moment $M^+_{s\ns_1,s\ns_2, \Bk}$.
However, when computing the valley magnetization at given fixed $\mu$, one must account for the Berry phase, which effectively modifies the density of states. The net result is that, in addition to summing the magnetic moment over the occupied states of the original spectrum, there is an additional contribution related to the Berry curvature.~\cite{Xiao2007vpg, Xiao2010bpe} Namely, ${\cal M} = {\cal M}\ns_{M} + {\cal M}_\Omega$, where
\begin{subequations}
\begin{equation}
  \label{eqn:def_M_M}
  {\cal M}\ns_{M} = g_{\rm s} \sum_{s\ns_2,s\ns_2} \,
  \int\!\dfrac{d^2\!k}{ (2\pi)^2} \> M\ns_\alpha\  \Theta\big(\mu-E\ns_\alpha\big) \,,
\end{equation}
$\alpha=\{s\ns_1,s\ns_2,\Bk\}$ is a composite index,
\begin{equation}
  {\cal M}\ns_{\Omega} = \dfrac{g_{\rm s}e}{\hbar c} \sum_{s\ns_1,s\ns_2} 
  \int\!\dfrac{d^2\!k}{ (2\pi)^2} \> \left( \mu - E\ns_\alpha \right)\ns_+ \> \Omega\ns_\alpha \,,
  \label{eqn:def_M_Omega}
\end{equation}
\end{subequations}
and $F\ns_+ \equiv F \,\Theta(F)$.
Thus, for $\mu > U$, in which case the occupied states of both conductions bands fill a circle, the integration limits are from $k = 0$ to $k = k\ns_{{\rm F},s\ns_2}$ where
\begin{equation}
  k\ns_{{\rm F},s\ns_2}=\dfrac{1}{\hbar v\ns_0}\sqrt{\mu^2+U^2-s\ns_2\sGamma^2(\mu)} \,.
\end{equation}
For $\Estar < \mu < U$, the occupied states fill an annulus in momentum space.
The limits on $k$ are from the inner radius $k\ns_{{\rm F},+}$ to the outer one $k\ns_{{\rm F},-}$.

Using the relation
\begin{equation}
  \Omega = \dfrac{1}{ 2\pi k} \dfrac{d\PhB}{d k}\,,
\end{equation}
which follows from Eq.~\eqref{eqn:Berry_curvature}, we reduce the expression for ${\cal M}\ns_\Omega$ to the integral over the Berry phase:
\begin{equation}
{\cal M}\ns_{\Omega} = \dfrac{g\ns_{\rm s} e}{\hbar c} \sum_{s\ns_1,s\ns_2} \,
\int \!\dfrac{d^2\!k}{ (2\pi)^2} \> \dfrac{|\vgr| }{ k} \, \dfrac{\PhB}{ 2\pi}  \  \Theta\big(\mu-E\ns_\alpha\big)\,.
\label{eqn:M_Omega}
\end{equation}
At this point we recall that the orbital magnetic moment $M^\pm_\alpha$ given by is related to the difference of the semiclassical and Berry phases, see Eq.~\eqref{eqn:orbital_m}.
As a result, the desired combination ${\cal M}\ns_M + {\cal M}\ns_\Omega$ is given by the integral over the semiclassical phase:
\begin{equation}
{\cal M} = \dfrac{g\ns_{\rm s} e}{\hbar c} \sum_{s\ns_1,s\ns_2} \,
\int \! \dfrac{d^2\!k}{(2\pi)^2} \> \dfrac{|\vgr| }{ k} \,
\dfrac{\Phc}{ 2\pi}  \  \Theta\big(\mu-E\ns_\alpha\big)\,.
\label{eqn:M_z_sc_phase}
\end{equation}
which can be evaluated in closed form.
The contribution from the two (partially occupied) conduction bands, using $g\ns_{\rm s}=2$, is
\begin{align}
  {\cal M}(\mu, U) &=-\dfrac{e\,U}{\pi\hbar c}+  \dfrac{e}{ \pi\hbar c} \cdot \dfrac{\Estar}{\gon} \label{eqn:M_z_cond_band}\\
  &\qquad\times\begin{cases}
    0 & \mu < \Estar\,, \\
    2 \sqrt{\mu^2 - E_\star^2} & \Estar \le \mu < U\,, \\
    \sqrt{\mu^2 - E_\star^2} + \dfrac{2 U \Estar}{\gon} & U \le \mu < E\ns_\diamond\,, \\
    \sqrt{4 U^2 + \gamma_1^2} & E\ns_\diamond \le \mu \,.\nonumber
  \end{cases}
\end{align}
Here the first term, which is linear in $U$, is due to the fully occupied valence bands, while the additional four possible contributions describe the contribution of the conduction bands.
Interestingly, once the higher energy band $s\ns_2 = +1$ becomes occupied, $\mu > E\ns_\diamond$, the total magnetization no longer depends on $\mu$, because of partial cancellation between the two conduction bands.

The function ${\cal M}(\mu,U)$ at $U= 80\,{\rm meV}$ is plotted in Fig.~\ref{fig:M_mu}(c).
Similar to the Hall conductivity, it has cusps at the energies where the
Fermi surface topology changes.
Specifically, for the first two of them we find
\begin{equation}
{\cal M}(\Estar) = -\dfrac{e \, U}{ \pi\hbar c} \,, \quad
{\cal M}(U) = -\dfrac{e\, U}{\pi\hbar c}\cdot
\dfrac{\gamma_1^2 }{ \gamma_1^2 + 4U^2 } \,.
\end{equation}

In Fig.~\ref{fig:M_mu}(c), the sign of ${\cal M}$ is negative.
However, this is unrelated to either paramagnetism or diamagnetism because the external magnetic field is assumed to be zero, in which case the $\Km$ valley makes an equal and opposite contribution to the total magnetization of the system.
Only the square of $M_\alpha$ contributes to the magnetic susceptibility:
\begin{equation}
\chi_P  = M_\alpha^2\, \nu = \frac14 g^2 \mu_\text{B}^2 \nu\,.
\label{eqn:chi_P}
\end{equation}
But $\chi_P$ is only one of the terms (known as the Pauli paramagnetism) which determine magnetic susceptibility.
As shown in previous work,~\cite{Safran1986sdo, Saito1986oso, Koshino2007odi, Koshino2010aom} the total susceptibility $\chi$ of BLG also contains the Landau diamagnetic term
\begin{equation}
\displaystyle
\chi_L = -\frac13 \left({m_\text{e}} / {m_\text{eff}}\right)^2 \mu_\text{B}^2 \nu\,,
\label{eqn:chi_L}
\end{equation}
as well as other contributions, which together generate a very complicated dependence of $\chi$ on $\mu$. (Here $\nu$ is the total electron density of states at the Fermi energy and $1 / m_{\rm eff} = d E_\alpha^2 / d k^2$ is the inverse effective mass.)

Concluding this section, we note that $M\ns_\alpha$, $\sigH(\mu)$, and ${\cal M}(\mu)$ in BLG were previously calculated numerically in Ref.~\onlinecite{Xiao2007vpg}.
Our analytic results for $M\ns_\alpha$ agree with that work.  (For the ease of comparison, a second axis is included in Fig.~\ref{fig:Mz}.) On the other hand, there are noticeable differences for $\sigH(\mu)$ and ${\cal M}(\mu)$.
Regarding $\sigH(\mu)$, we suspect that the authors of Ref.~\onlinecite{Xiao2007vpg} included the effect of impurity scattering in the form of the side-jump, which we ignore.
The plot of ${\cal M}(\mu)$ presented in Ref.~\onlinecite{Xiao2007vpg} lacks the cusps that should be there due to the changes in the Fermi surface topology, see our Fig.~\ref{fig:M_mu}(c).

%%%%%%%%%%%%%%%%%%%%%%%%%%%%%%%%%%%%%%%%%%%%%%%%%%%%%%%%%%%%%%%%%%%%%%%%%%%%%%%
\section{Discussion and conclusions}
\label{sec:conclusion}
%%%%%%%%%%%%%%%%%%%%%%%%%%%%%%%%%%%%%%%%%%%%%%%%%%%%%%%%%%%%%%%%%%%%%%%%%%%%%%%
In this paper we have presented a quasiclassical Landau quantization procedure which includes both the Berry phase and the magnetoelectric effects on the band structure.
This method provides an intuitive picture of the Landau level dispersion and several other measurable properties of biased BLG.
In some cases, we have been able to derive analytic expressions for the Landau level energies; we also discussed how they may be computed numerically.

Our results are applicable in the analysis of a number of experiments which probe transport and thermodynamic properties of BLG, including cyclotron resonance, activated conductivity, charge compressibility, and magnetization.
Of course, a more realistic calculation of these quantities should also include interaction effects.
The self-consistent mean field approximation for BLG has been addressed in several published works, but generally such treatments have neglected exchange and correlation effects, which were considered in
Refs.~\onlinecite{Barlas2008ill, Kusminskiy2009eei, Abergel2009lrc, Abanin2009c2e} and shown to give as much as a $\sim 30\%$ correction to the mean field (Hartree) approximation, similar to the case in two-dimensional (2D) electron systems in semiconductors~\cite{Kallin1984efa, *[{for discussion of large $\nu$, see also }][ and references therein]Fogler1995cos}\nocite{Fogler1995cos} % DO NOT REMOVE THIS NOCITE, It doesn't affect anything, but some of the script I use (MZ).
Currently, experimental results for the Landau level energies from the cyclotron resonance~\cite{Henriksen2008cri} and the charge compressibility studies~\cite{Henriksen2010mot} can be fitted to the theory if undetermined variables ($U$, for example) are treated as adjustable parameters. 
Incorporating all major experimentally relevant ingredients -- Hartree, exchange, and disorder contributions --- into the same calculation would be a more stringent test of the theory.

Although the Landau level dispersion and therefore Landau level crossing points cannot yet be calculated with a high degree of accuracy, phenomena that may be observed at such points are quite interesting.
Indeed, crossing of Landau levels has been previously studied~\cite{Kallin1984efa, Falko1993ccs, Jungwirth1998mai, Jungwirth2000pac, Chalker2002qhf, Wang2002mea, Rezayi2003eds, Koch1993sll, Cho1998hst, Piazza1999fop, DePoortere2000rsa, Eom2000qhf, Zeitler2001mai, dePoortere2003cri, Pan2001hat, ZhangX2005mpw, ZhangX2006lla, Gusynin2006tod, Fischer2007teo} in the context of the quantum Hall effect in conventional 2D systems.
In those systems, the crossings are between Landau levels of different subbands or between spin-split levels of different Landau levels of the same subband.
Near the crossing the energy gap vanishes, and so a spike in conductance is expected.
In the quantum Hall effect conditions, this is simultaneously a spike in resistance.
In experiments, such spikes have been observed to be hysteretic.
Sometimes, they were also accompanied by a spatial anisotropy of the transport.
The leading theoretical explanation~\cite{Jungwirth1998mai} attributes these phenomena to quantum Hall ferromagnetism (QHF).
Namely, when two
Landau level are nearly degenerate and the chemical potential is close to the crossing energy, the occupation of the Landau levels are modeled as two states pseudo-spin system.
Depending on the nature of the crossing, QHF can be of  either easy-axis or easy-plane type.
In the former case, one expects formation of domains whose collective dynamics can in principle generate both hysteresis and anisotropy.
The BLG appears a promising system to study
QHF because of its high tunability and a rich pattern of level crossings we discussed in the paper.

We are particularly grateful to F.~Guinea for valuable interactions in the early stages of this work.
We also thank
V.~Fal'ko and Q.~Niu for discussions.
This work is supported by the NSF under Grant DMR-0706654 and ASC UCSD Grant
RG209G (LMZ and MMF).
DPA is grateful to the Aspen Center for Physics, where some of this work was performed.

\appendix
%%%%%%%%%%%%%%%%%%%%%%%%%%%%%%%%%%%%%%%%%%%%%%%%%%%%%%%%%%%%%%%%%%%%%%%%%%%%%%%
\section{Low-energy theory of BLG}
\label{sec:low_energy_H_Eff}
%%%%%%%%%%%%%%%%%%%%%%%%%%%%%%%%%%%%%%%%%%%%%%%%%%%%%%%%%%%%%%%%%%%%%%%%%%%%%%%

In this section we derive the low-energy of BLG by the standard method of canonical transformation.
Our results are in a good agreement with previous work.~\cite{McCann2006lld, McCann2007eib, Nilsson2008epo} Some minor discrepancies can be attributed to typographic errors therein or differences in notations.

We begin with the Hamiltonian of Eq.~\eqref{Hbilayer}.
The bilayer's electronic structure has four bands.
When $\Bq=\BK^\pm$, the two central levels lie at $E=\pm U$.
For $|S\ns_\Bq|\ll 1$, where $S_\Bq$ is given in Eq.~\eqref{eqn:wallace}, we can derive an effective $2\times 2$ Hamiltonian by writing
\begin{equation}
H\ns_{\BK+\Bk} = H^0 + V\,, 
\label{eqn:V}
\end{equation}
where $H^0=H(\BK^\pm)$ contains terms dependent on $\gon$, $\Dpr$, and $U$, and $V$ contains the $\gze S_\Bq$, $\gth S_\Bq$, and $\gfo S_\Bq$ terms. (To lighten notations, the subscript $\Bq$ is dropped in the following.)

The unperturbed Hamiltonian $H^0$ has levels $E^0_{1,4} = \mp\sqrt{\gamma_1^2 + U^2}$ and $E^0_{2,3} = \Dpr\mp U$.
The eigenfunctions $\ket{\psi\ns_j}$ are the column vectors of the matrix
\begin{equation}
  \Psi=
  \begin{pmatrix} 0 & 1 & 0 & 0 \\
    \cos(\theta/2) & 0 & 0 & \sin(\theta/2) \\
    -\sin(\theta/2) & 0 & 0 & \cos(\theta/2) \\ 
    0 & 0 & 1 & 0
  \end{pmatrix}\,,
\end{equation}
where $\tan\theta=\gon/U$.
Eliminating the high energy  subspace spanned by $\ket{\psi\ns_{1,4}}$ by unitary transformation
\begin{equation}
  \Htil=e^{i Q} H e^{-i Q}\,,
\end{equation}
we obtain the effective $2\times 2$ Hamiltonian
\begin{align}
  \Htil\ns_{nn'}&=E^0_n\,\delta\ns_{nn'} + V\ns_{nn'} \nonumber\\
  &\qquad +\half\sum_a \bigg(\dfrac{1}{ E^0_n-E^0_a} + \dfrac{1}{E^0_{n'} - E^0_a}\bigg)\,V\ns_{n\ns a}\,V\ns_{an'}\,,
\end{align}
up to terms of order $V^2$.
Here $n,n'\in\{2,3\}$ are labels for the low energy subspace, while $a\in\{1,4\}$ labels the high energy subspace, and $V\ns_{ij}=\expect{\psi\ns_i}{V}{\psi\ns_j}$.
The matrix elements of $Q$ are given by
\begin{equation}
\begin{split}
Q\ns_{na} &= i\, \dfrac{V_{na}}{E^0_a - E^0_n}
 + i \sum_{n'} \frac{V_{nn'} V_{n'a}}
                    {(E^0_a - E^0_n)(E^0_a - E^0_{n'})}\\
&\qquad -i\sum_{a'} \frac{V_{na'} V_{a'a}}
                         {(E^0_a - E^0_n)(E^0_{a'} - E^0_{n})}
                          + o(V^2)\,,
\end{split}
\end{equation}
with $Q\ns_{an} = (Q\ns_{an})\yd$.

Proceeding in this manner, we obtain the $2\times 2$ block for the inner bands,
\begin{equation}
{\widetilde H}=
\begin{pmatrix}
\varepsilon\ns_0 - U +\omega & \xi \\
\xi^* & \varepsilon\ns_0 + U - \omega
\end{pmatrix}\,,
\label{effectiveham}
\end{equation}
where, to lowest order in $U$ and $\Dpr$,
\begin{align}
\varepsilon\ns_0 &=\Bigg(\dfrac{\gze\gfo}{\gon} + \dfrac{(\gamma_0^2+\gamma_4^2)\,\Dpr}{ 2\gamma_1^2}\Bigg) \big\{S,S\yd\big\}
+ \dfrac{U\gamma_0^2}{\gamma_1^2}\, \big[S,S\yd\big]\,,
\\
\omega &= \dfrac{U\gamma_0^2}{\gamma_1^2}\,\big\{S,S\yd\big\} +
\Bigg(\dfrac{\gze\gfo}{\gon} + \dfrac{(\gamma_0^2+\gamma_4^2)\,\Dpr}{ 2\gamma_1^2}\Bigg) \big[S,S\yd\big]\,,
\\
\xi &=\gth\,S\yd - \Bigg(\dfrac{\gamma_0^2+\gamma_4^2}{ \gon} + \dfrac{2\gze\gfo\Dpr}{\gamma_1^2}\Bigg) S^2\,.
\end{align} 
Anticipating the introduction of an external magnetic field, we have allowed for the possibility that $S$ and $S\yd$ do not commute.
Recognizing that $\gfo/\gze=0.05 \ll 1$, it is permissible to drop the terms of order $\gamma_4^2$ and $\gfo\Dpr$, in which case
\begin{align}
\ve\ns_0 &=\dfrac{\gamma_0^2\,\Dtil}{ 2\gamma_1^2}\, \big\{S,S\yd\big\}
+ \dfrac{U\gamma_0^2}{\gamma_1^2}\, \big[S,S\yd\big]\,, \\
\omega &= \dfrac{U\gamma_0^2}{\gamma_1^2}\,\big\{S,S\yd\big\} +
\dfrac{\gamma_0^2\,\Dtil}{ 2\gamma_1^2}\, \big[S,S\yd\big]\,, \\
\xi &= \gth\,S\yd - \dfrac{\gamma_0^2}{ \gon}  S^2\,,
\end{align}
leading to Eq.~\eqref{eqn:H_eff_full}.
Our results agree with those of Ref.~\onlinecite{McCann2006lld} if $\Dpr$ and $\gfo$ are set to zero. 

For $B = 0$, in the vicinity of the $\BK^\pm$ points, the four bands disperse as shown in Fig.~\ref{fig:BLG_E_bands}.
The two central bands, which comprise the low energy sector, are separated by $2U$ at $\Bk=0$.
Their dispersion is described by the effective Hamiltonian of Eq.~\eqref{effectiveham}.
One finds that for $\Bk = k\,\xhat$ the central bands have a characteristic double hump (or Mexican hat) shape provided $U > 2\gon\gth/\gze\approx 80\,$meV.

It is convenient to write ${\widetilde H} = \ve\ns_0 + \BCB(\Bk)\,{\mib\sigma}$, where $\CB_z = \omega$, $\CB_x - i\CB_y = \xi$, and ${\mib\sigma}$ is the vector of Pauli matrices.
When the actual magnetic field $B$ vanishes, 
\begin{gather}
\ve\ns_0 = \dfrac{\Dtil\,\ve_k^2}{\gamma_1^2} \,,\quad \omega = \dfrac{2\ve_k^2}{\gamma_1^2}\,U\,,\\
\xi = \dfrac{\gth}{\gze}\,\ve\ns_k\,e^{-i\vphi} - \dfrac{\ve_k^2}{\gon}\,e^{+2i\vphi}\,,
\end{gather}
where $\ve\ns_k = \hbar v_0 k$, the origin in $\Bk$-space is taken as one of the $K^\pm$ points, and $\vphi = \tan^{-1}(k\ns_y / k\ns_x)$ is the corresponding polar angle.
The eigenvalues are
\begin{equation}
\Etil\ns_{s\ns_1,-,\Bk}=\ve\ns_0 + s\ns_1 |\BCB(\Bk)|\,.
\end{equation}

Let us now discuss the Berry phase.
Semiclassically, in the presence of a weak magnetic field, the wavevector $\Bk$ evolves in time according to Eq.~\eqref{eqn:kdot}:
\begin{equation*}
\dot{\Bk} = \omega_{\rm c} \, \zhat \times \Bk \quad,\quad
\omega_{\rm c} \equiv \dfrac{2 \pi }{ T} \sign(\vgr)\,.
\end{equation*}
If we can neglect $\gth$, then the trajectory the pseudospin traces on the Bloch sphere winds {\it twice} for every cycle of $\Bk$, owing to the $e^{2i\vphi}$ factor in $\CB_x-i\CB_y=\xi$.
Therefore, the accumulated Berry phase is equal to $2 \times \half = 1$ times the solid angle traced by vector $\BCB(\Bk)$.
Actually, the Berry phase is defined modulo $2\pi$.
To be consistent with the earlier choice of the overall phase factor of the basis state~\eqref{eqn:psi_B_0}, we need to subtract $2 \pi$ from the solid angle.
The result is
\begin{equation}
\Phi_\text{B}^\prime = 2\pi \bigg( 1 + \dfrac{\CB_z}{ |\BCB|}\bigg) - 2\pi
= \dfrac{2\pi U}
         {E\ns_{s\ns_1,-, k}}
         \bigg( \dfrac{2\ve_k^2}{\gamma_1^2}-1\bigg)\,.
\end{equation}
However, it differs from our earlier Eq.~\eqref{eqn:Berry_phase_final} for the Berry phase in BLG.
The discrepancy arises due to the canonical transformation by which we obtain the wavefunction in the new basis: $\ket{\psi^{\prime}} = e^{-i Q} \ket{\psi}$:
\begin{equation}
\begin{split}
 \delta \Phi &\equiv \PhB - \Phi_{\rm B}^\prime
             = -{2 \pi i} \expect{\psi'} { e^{i Q} \,
               \partial_{\varphi}\, e^{-i Q} } {\psi'} \\
             &= {2\pi} \dfrac{\ve_k^2}{\gamma_1^2}
                       \dfrac{\CB_z}{ |\BCB|}
             \simeq \dfrac{\ve_k^2}{\gamma_1^2}\,
                    \dfrac{2\pi U}{E\ns_{s\ns_1,-, k}} \,.
\label{eqn:delta_Phi}
\end{split}
\end{equation}
The combined phase $\PhB = \Phi_{\rm B}^\prime + \delta \Phi$ is in agreement with the four-band expression Eq.~\eqref{eqn:Berry_phase_final}, to within the accuracy of this calculation.

%%%%%%%%%%%%%%%%%%%%%%%%%%%%%%%%%%%%%%%%%%%%%%%%%%%%%%%%%
\begin{figure}[bt]
\centering
\includegraphics[width=3.0in]{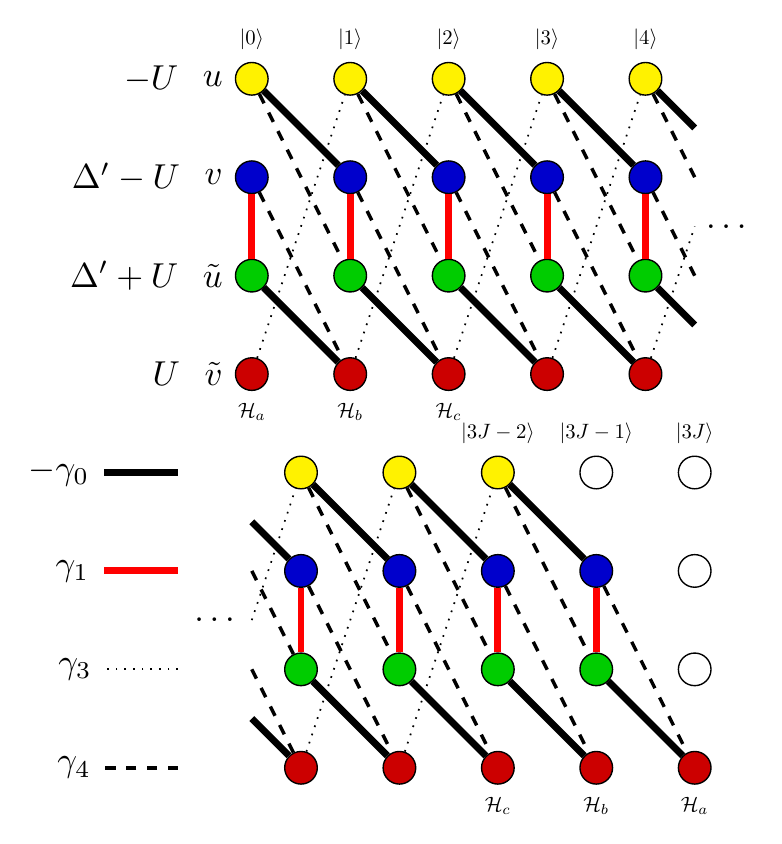}
\caption{(Color online) Sketch of the structure of the magnetic bilayer Hamiltonian, showing nonzero matrix elements as links.
Each link between orbitals in column $n$ and column $n+1$ is multiplied by a factor $\sqrt{n+1}\,(\omega_0 / \gamma_0)$.
The diagonal entries in the Hamiltonian for each orbital are given at the upper left.
When $\gamma_3=0$, the Hamiltonian breaks up into a direct sum of $4\times 4$ blocks.
\label{fig:magham}}
\end{figure}
%%%%%%%%%%%%%%%%%%%%%%%%%%%%%%%%%%%%%%%%%%%%%%%%%%%%%%%%%

Finally, we can go beyond the semiclassical approximation, obtaining the effective Hamiltonian in Eq.~\eqref{eqn:HeffB}.
If $\gth$ is neglected, each pseudospin component Landau level is connected to a unique mate, and the Hamiltonian breaks up into a direct sum of $2\times 2$ blocks, given by Eq.~\eqref{Htilp} ($n=-1$ and $n=0$ are special cases where $\Htil$ reduces to a scalar).

%%%%%%%%%%%%%%%%%%%%%%%%%%%%%%%%%%%%%%%%%%%%%%%%%%%%%%%%%%%%%%%%%%%%%%%%%%%%%%%
\section{Matrix representation of the Hamiltonian in a finite magnetic field}
\label{sec:BLG_H_n}
%%%%%%%%%%%%%%%%%%%%%%%%%%%%%%%%%%%%%%%%%%%%%%%%%%%%%%%%%%%%%%%%%%%%%%%%%%%%%%%

In the presence of a magnetic field, the full Hamiltonian $\CH^+$ in the $K^+$ valley is given by Eq.~\eqref{HpB},
\begin{equation*}
\CH^+=\begin{pmatrix} -U & -\omo\,a & \eta\ns_4\,\omo\,a & \eta\ns_3\,\omo\,a\yd \\
-\omo\,a\yd & -U+\Dpr  & \gon & \eta\ns_4\,\omo\,a \\
\eta\ns_4\,\omo\,a\yd & \gon & U+ \Dpr  & -\omo\,a \\
\eta\ns_3\,\omo\,a & \eta\ns_4\,\omo\,a\yd & -\omo\,a\yd & U \end{pmatrix}\,,
\end{equation*}
where $a$ and $a\yd$ are Landau level lowering and raising operators, respectively.
In the occupation number basis $\ket{n}$, the matrix elements of $\CH^+$ can be understood pictorially, by referring to Fig.~\ref{fig:magham}.
Writing the general wavefunction as
\begin{equation}
\ket{\Psi}=\sum_{n=0}^\infty \begin{pmatrix} u\ns_n \ket{n},\, v\ns_n \ket{n},\, \util\ns_n \ket{n},\, \vtil\ns_n \ket{n} \end{pmatrix}^{\! \textsf T},
\end{equation}
the links in Fig.~\ref{fig:magham} indicate matrix elements between the various components $\{u\ns_n,v\ns_n,\util\ns_n,\vtil\ns_n\}$.

\begin{figure}[bt]
  \begin{center}
    \includegraphics[width=3.2in]{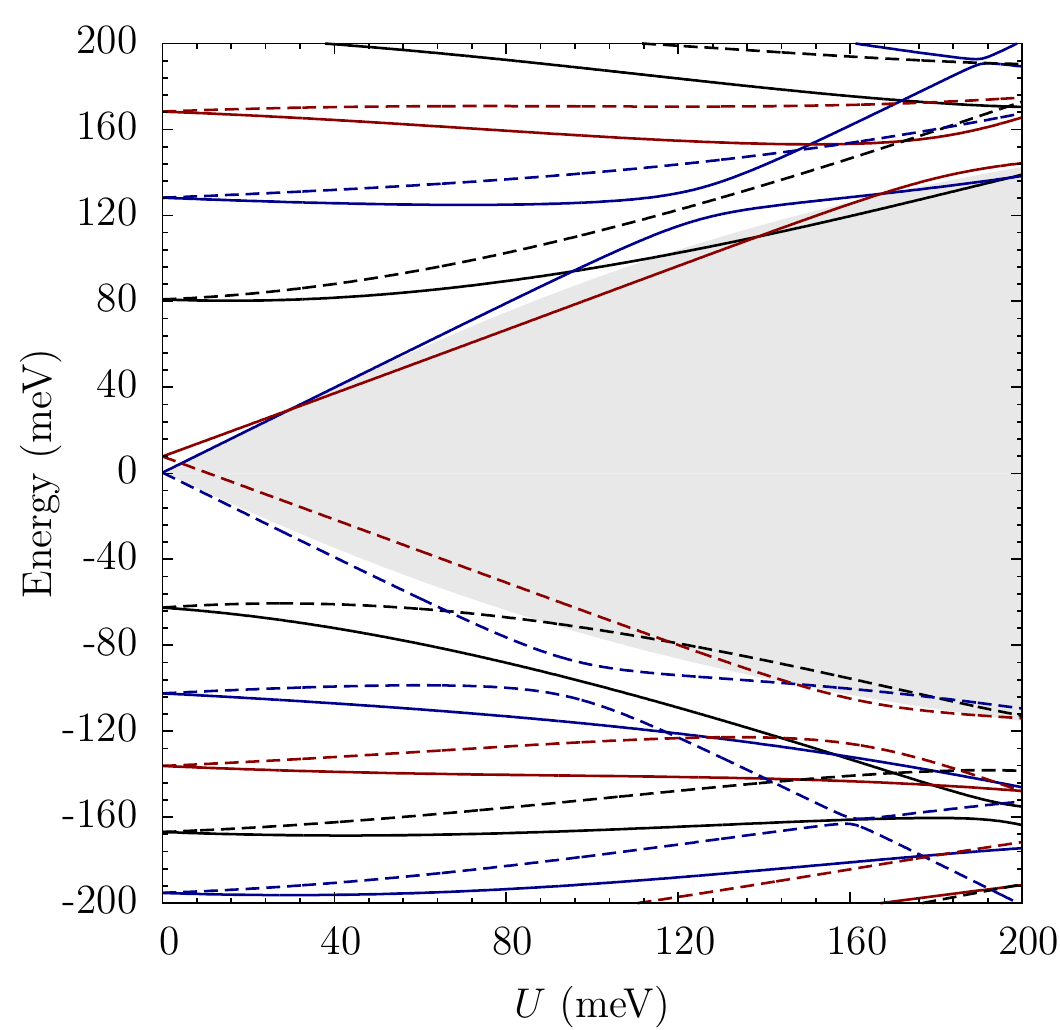}
  \end{center}
  \caption{(Color online) Landau level energies {\it vs.} interlayer bias $U$ for a field value $B=20\,$T.
  Solid lines correspond to the $K^+$ valley and broken lines to the $K^-$ valley.
  The color and line type are as in Fig.~\ref{fig:vspecb}.
  The shaded area indicates the energy gap at $B = 0$.}
  \label{fig:vspeca}
\end{figure}

One finds that  $\CH=\CH\ns_{\rm a}\oplus\CH\ns_{\rm b}\oplus\CH\ns_{\rm c}$ can be written as a direct sum of three terms.
In evaluating the spectrum numerically, we truncate $\CH_{\rm a,b,c}$ at a high Landau level index, as shown in the figure.
Typically we chose a maximum index of $n\ns_{\rm max}\approx 300$, checking that spectrum did not vary significantly as the upper index cutoff was further increased.
This feature is most evident at high fields, such as in Fig.~\ref{fig:vspeca}, where we have taken $B=20\,$T.
The spectrum of $\CH_{\rm a}$ is shown in black, that of $\CH_{\rm b}$ in red, and that of $\CH_{\rm c}$ in blue.
Solid lines correspond to the $K^+$ valley and broken lines to the $K^-$ valley.
One sees in the figure that curves of the same color and line type cannot cross at an accidental degeneracy.

Frequently in this paper we have ignored the SWMc parameter $\gth$, setting it to zero,  In this approximation, as can be seen from Fig.~\ref{fig:magham}, the occupation number space Hamiltonian further resolves itself into a direct sum of $4\times 4$ blocks, given by the expression in Eq.~\eqref{Hnblock}, which connect $\{u\ns_{n-1}\,,\,v\ns_n\,,\,\util\ns_n\,,\,v\ns_{n+1}\}$ for each $n$.  (There is also a remaining $1\times 1$ and $3\times 3$ block associated with the indices $n=0$ and $n=1$.)

%%%%%%%%%%%%%%%%%%%%%%%%%%%%%%%%%%%%%%%%%%%%%%%%%%%%%%%%%
%%%%%%%%%%%%%%%%%%%%%%%%%%%%%%%%%%%%%%%%%%%%%%%%%%%%%%%%%

In addition to eigenvalues, we also calculated the eigenfunctions, one of which is shown in Fig.~\ref{fig:pockets}.
To do so we chose the symmetric gauge, where the Bloch wavefunctions of $\ket{m}$ oscillator states are given by
\begin{equation}
\braket{\Bk}{n}=
\frac{\ell_{\!B}^{m + 1}}{\sqrt{m!}}\,
\left(\frac{k_x - i k_y}{\sqrt{2}}\right)^{\!\!m}
 e^{-k^2 \ell_{\!B}^2 / 4}\,.
\label{eqn:ket_m_Bloch}
\end{equation}
These basis states were weighted with the coefficients obtained from diagonalizing the Hamiltonian matrix and then summed over all components (both $n$ and the sublattice index).

From these calculations we concluded that the effect of $\gth$ diminishes as $B$ increases, as was previously observed in Ref.~\onlinecite{McCann2006lld} The semiclassical argument that explains this behavior was given in Sec.~\ref{sec:trigonal}.
Here we mention another reasoning,~\cite{McCann2006lld} which is based on the usual perturbation theory.

The leading-order correction to the energies due to $\gth$ is approximately $\omega_0^2/\Delta E$, where $\Delta E \approx \hbar |\vgr| / (k\ell_B^2)$ is the Landau level spacing.
Therefore, the relative magnitude of this energy shift is small provided
\begin{equation}
\gth\ll \dfrac{\gze}{ k\ellB}\,.
\label{eqn:gamma_3_condition}
\end{equation}
At $U = 0$ and $\ve\ns_k \ll \gon$ this inequality gives~\cite{McCann2006lld} $\omega_0 \gg \gth\gon/\gze$, which is roughly consistent with the threshold $B \sim 1\,$T where the effect of $\gth$ is observed to become insignificant in the numerical calculations.
On the other hand, at finite $U$ and near the bottom of the Mexican hat, where $\vgr = 0$, the expression on the right-hand side of
Eq.~\eqref{eqn:gamma_3_condition} diverges.
This implies that the effect of $\gamma\ns_3$ is larger and persists to higher $B$.
This is also consistent with the numerics, see Sec.~\ref{sec:trigonal}.

%%%%%%%%%%%%%%%%%%%%%%%%%%%%%%%%%%%%%%%%%%%%%%%%%%%%%%%%%%%%%%%%%%%%%%%%%%%%%%%
%Merlin.mbs v4.21 2009-07-09.
%
\end{document}